\RequirePackage{fix-cm}
\documentclass[smallextended]{svjour3}
\smartqed
\usepackage{appendix}
\usepackage{amsmath}
\usepackage{graphicx}
\usepackage{longtable}
\usepackage{lineno} 
\usepackage{etoolbox}

\newtoggle{linenumbersenabled}
\toggletrue{linenumbersenabled}

\AtBeginEnvironment{align}{\iftoggle{linenumbersenabled}{\linenumbers}{}}
\AtEndEnvironment{align}{\iftoggle{linenumbersenabled}{\nolinenumbers}{}}

\usepackage{array}
\usepackage{longtable}
\usepackage{tikz}
\usepackage{natbib}
\setcitestyle{authoryear,open={(},close={)}} 
\usepackage{float,mwe,amssymb,bm}
\newcommand\numberthis{\addtocounter{equation}{1}\tag{\theequation}}
\newcommand{\mean}[1]{\left\langle#1\right\rangle}

\newcommand\T[1]{\hat{\mathrm{#1}}}
\newcommand\To[1]{\hat{\mathrm{#1}}^{\text{outer}}}
\newcommand\Ti[1]{\hat{\mathrm{#1}}^{\text{inner}}}

\usepackage{hyperref}

\definecolor{color1}{RGB}{255, 0, 0}         
\definecolor{color2}{RGB}{119, 172, 48}      
\definecolor{color3}{RGB}{0, 0, 0}           
\definecolor{color4}{RGB}{255, 0, 255}       
\definecolor{color5}{RGB}{0, 255, 255}       
\definecolor{color6}{RGB}{126, 47, 142}      
\definecolor{color7}{RGB}{0, 0, 255}         
\definecolor{color8}{RGB}{237,177, 32}      
\definecolor{MATLAB_green}{RGB}{0,255, 0}      

\newcommand{\CNBL}{\raisebox{2pt}{\tikz{\draw[-,color1,solid,line width = 1.0pt](0,0) -- (5mm,0);}}}
\newcommand{\SBLa}{\raisebox{2pt}{\tikz{\draw[-,color2,solid,line width = 1.0pt](0,0) -- (5mm,0);}}}
\newcommand{\SBLb}{\raisebox{2pt}{\tikz{\draw[-,color3,solid,line width = 1.0pt](0,0) -- (5mm,0);}}}
\newcommand{\SBLc}{\raisebox{2pt}{\tikz{\draw[-,color4,solid,line width = 1.0pt](0,0) -- (5mm,0);}}}
\newcommand{\SBLd}{\raisebox{2pt}{\tikz{\draw[-,color5,solid,line width = 1.0pt](0,0) -- (5mm,0);}}}
\newcommand{\SBLe}{\raisebox{2pt}{\tikz{\draw[-,color6,solid,line width = 1.0pt](0,0) -- (5mm,0);}}}
\newcommand{\SBLf}{\raisebox{2pt}{\tikz{\draw[-,color7,solid,line width = 1.0pt](0,0) -- (5mm,0);}}}
\newcommand{\model}{\raisebox{0pt}{\tikz{\draw[-, color8, solid, line width = 1.0pt](0,0) -- (5mm,0);
\draw[color8,line width = 1.0pt] (2.5mm, 0) circle (2pt);}}}

\newcommand{\Toyzmodelfa}{\raisebox{0pt}{\tikz{\draw[-, color1, dash dot, line width = 1.0pt](0,0) -- (5mm,0);
\draw[color1,line width = 1.0pt] (2.5mm, 0) circle (2pt);}}}
\newcommand{\Toyzmodelfb}{\raisebox{0pt}{\tikz{\draw[-, color2, dash dot, line width = 1.0pt](0,0) -- (5mm,0);
\draw[color2,line width = 1.0pt] (2.5mm, 0) circle (2pt);}}}
\newcommand{\Toyzmodelfc}{\raisebox{0pt}{\tikz{\draw[-, color3, dash dot, line width = 1.0pt](0,0) -- (5mm,0);
\draw[color3,line width = 1.0pt] (2.5mm, 0) circle (2pt);}}}
\newcommand{\Toyzmodelfd}{\raisebox{0pt}{\tikz{\draw[-, color4, dash dot, line width = 1.0pt](0,0) -- (5mm,0);
\draw[color4,line width = 1.0pt] (2.5mm, 0) circle (2pt);}}}
\newcommand{\Toyzmodelfe}{\raisebox{0pt}{\tikz{\draw[-, color5, dash dot, line width = 1.0pt](0,0) -- (5mm,0);
\draw[color5,line width = 1.0pt] (2.5mm, 0) circle (2pt);}}}
\newcommand{\Toyzmodelff}{\raisebox{0pt}{\tikz{\draw[-, color6, dash dot, line width = 1.0pt](0,0) -- (5mm,0);
\draw[color6,line width = 1.0pt] (2.5mm, 0) circle (2pt);}}}
\newcommand{\Toyzmodelfg}{\raisebox{0pt}{\tikz{\draw[-, color7, dash dot, line width = 1.0pt](0,0) -- (5mm,0);
\draw[color7,line width = 1.0pt] (2.5mm, 0) circle (2pt);}}}

\newcommand{\Ttotmodel}{\raisebox{0pt}{\tikz{
  \draw[-, color8, solid, line width = 1.0pt](0,0) -- (5mm,0);
  \draw[color8,fill=none, line width=1.0pt,rotate=90] (-0.70mm,-2.00mm) -- ++(-60:1.45mm) -- ++(60:1.45mm) -- cycle;
}}}

\newcommand{\Txzmodel}{\raisebox{0pt}{\tikz{
  \draw[-, color8, solid, line width = 1.0pt](0,0) -- (5mm,0);
  \draw[color8,fill=none, line width=1.0pt,rotate=0] 
  (2.5mm-0.65mm,+0.650mm) -- (2.5mm+0.65mm,+0.650mm) -- (2.5mm+0.65mm,-0.650mm) -- 
  (2.5mm-0.65mm,-0.650mm) -- cycle;
}}}
\newcommand{\Txzmodelfa}{\raisebox{0pt}{\tikz{
  \draw[-, color1, dashed, line width = 1.0pt](0,0) -- (5mm,0);
  \draw[color1,fill=none, line width=1.0pt,rotate=0] 
  (2.5mm-0.65mm,+0.650mm) -- (2.5mm+0.65mm,+0.650mm) -- (2.5mm+0.65mm,-0.650mm) -- 
  (2.5mm-0.65mm,-0.650mm) -- cycle;
}}}
\newcommand{\Txzmodelfb}{\raisebox{0pt}{\tikz{
  \draw[-, color2, dashed, line width = 1.0pt](0,0) -- (5mm,0);
  \draw[color2,fill=none, line width=1.0pt,rotate=0] 
  (2.5mm-0.65mm,+0.650mm) -- (2.5mm+0.65mm,+0.650mm) --(2.5mm+0.65mm,-0.650mm) -- (2.5mm-0.65mm,-0.650mm) -- cycle;
}}}
\newcommand{\Txzmodelfc}{\raisebox{0pt}{\tikz{
  \draw[-, color3, dashed, line width = 1.0pt](0,0) -- (5mm,0);
  \draw[color3,fill=none, line width=1.0pt,rotate=0] 
  (2.5mm-0.65mm,+0.650mm) -- (2.5mm+0.65mm,+0.650mm) --(2.5mm+0.65mm,-0.650mm) -- (2.5mm-0.65mm,-0.650mm) -- cycle;
}}}
\newcommand{\Txzmodelfd}{\raisebox{0pt}{\tikz{
  \draw[-, color4, dashed, line width = 1.0pt](0,0) -- (5mm,0);
  \draw[color4,fill=none, line width=1.0pt,rotate=0] 
  (2.5mm-0.65mm,+0.650mm) -- (2.5mm+0.65mm,+0.650mm) --(2.5mm+0.65mm,-0.650mm) -- (2.5mm-0.65mm,-0.650mm) -- cycle;
}}}
\newcommand{\Txzmodelfe}{\raisebox{0pt}{\tikz{
  \draw[-, color5, dashed, line width = 1.0pt](0,0) -- (5mm,0);
  \draw[color5,fill=none, line width=1.0pt,rotate=0] 
  (2.5mm-0.65mm,+0.650mm) -- (2.5mm+0.65mm,+0.650mm) --(2.5mm+0.65mm,-0.650mm) -- (2.5mm-0.65mm,-0.650mm) -- cycle;
}}}
\newcommand{\Txzmodelff}{\raisebox{0pt}{\tikz{
  \draw[-, color6, dashed, line width = 1.0pt](0,0) -- (5mm,0);
  \draw[color6,fill=none, line width=1.0pt,rotate=0] 
  (2.5mm-0.65mm,+0.650mm) -- (2.5mm+0.65mm,+0.650mm) --(2.5mm+0.65mm,-0.650mm) -- (2.5mm-0.65mm,-0.650mm) -- cycle;
}}}
\newcommand{\Txzmodelfg}{\raisebox{0pt}{\tikz{
  \draw[-, color7, dashed, line width = 1.0pt](0,0) -- (5mm,0);
  \draw[color7,fill=none, line width=1.0pt,rotate=0] 
  (2.5mm-0.65mm,+0.650mm) -- (2.5mm+0.65mm,+0.650mm) --(2.5mm+0.65mm,-0.650mm) -- (2.5mm-0.65mm,-0.650mm) -- cycle;
}}}

\newcommand{\Tiyzmodel}{\raisebox{0pt}{\tikz{
  \draw[-, color8, solid, line width = 1.0pt](0,0) -- (5mm,0);
  \draw[color8,fill=none, line width=1.0pt] (1.75mm,0.60mm) -- ++(-60:1.45mm) -- ++(60:1.45mm) -- cycle;
}}}
\newcommand{\Tiyzmodelfa}{\raisebox{0pt}{\tikz{
  \draw[-, color1, dash dot, line width = 1.0pt](0,0) -- (5mm,0);
  \draw[color1,fill=none, line width=1.0pt] (1.75mm,0.60mm) -- ++(-60:1.45mm) -- ++(60:1.45mm) -- cycle;
}}}
\newcommand{\Tiyzmodelfb}{\raisebox{0pt}{\tikz{
  \draw[-, color2, dash dot, line width = 1.0pt](0,0) -- (5mm,0);
  \draw[color2,fill=none, line width=1.0pt] (1.75mm,0.60mm) -- ++(-60:1.45mm) -- ++(60:1.45mm) -- cycle;
}}}
\newcommand{\Tiyzmodelfc}{\raisebox{0pt}{\tikz{
  \draw[-, color3, dash dot, line width = 1.0pt](0,0) -- (5mm,0);
  \draw[color3,fill=none, line width=1.0pt] (1.75mm,0.60mm) -- ++(-60:1.45mm) -- ++(60:1.45mm) -- cycle;
}}}
\newcommand{\Tiyzmodelfd}{\raisebox{0pt}{\tikz{
  \draw[-, color4, dash dot, line width = 1.0pt](0,0) -- (5mm,0);
  \draw[color4,fill=none, line width=1.0pt] (1.75mm,0.60mm) -- ++(-60:1.45mm) -- ++(60:1.45mm) -- cycle;
}}}
\newcommand{\Tiyzmodelfe}{\raisebox{0pt}{\tikz{
  \draw[-, color5, dash dot, line width = 1.0pt](0,0) -- (5mm,0);
  \draw[color5,fill=none, line width=1.0pt] (1.75mm,0.60mm) -- ++(-60:1.45mm) -- ++(60:1.45mm) -- cycle;
}}}
\newcommand{\Tiyzmodelff}{\raisebox{0pt}{\tikz{
  \draw[-, color6, dash dot, line width = 1.0pt](0,0) -- (5mm,0);
  \draw[color6,fill=none, line width=1.0pt] (1.75mm,0.60mm) -- ++(-60:1.45mm) -- ++(60:1.45mm) -- cycle;
}}}
\newcommand{\Tiyzmodelfg}{\raisebox{0pt}{\tikz{
  \draw[-, color7, dash dot, line width = 1.0pt](0,0) -- (5mm,0);
  \draw[color7,fill=none, line width=1.0pt] (1.75mm,0.60mm) -- ++(-60:1.45mm) -- ++(60:1.45mm) -- cycle;
}}}

\newcommand{\Umodel}{\raisebox{0pt}{\tikz{
  \draw[-, black, dashed, line width = 1pt](0,0) -- (5mm,0);
  \draw[black,fill=black, line width=0.5pt] (1.75mm,0.60mm) -- ++(-60:1.45mm) -- ++(60:1.45mm) -- cycle;
}}}
\newcommand{\UMOST}{\raisebox{2pt}{\tikz{
  \draw[-, color2, dash dot, line width = 1pt](0,0) -- (5mm,0);
}}}

\newcommand{\Vmodela}{\raisebox{0pt}{\tikz{\draw[-, color1, solid, line width = 1.0pt](0,0) -- (5mm,0);
\draw[color1,fill,line width = 1.0pt] (2.5mm, 0) circle (2pt);}}}
\newcommand{\Vmodelb}{\raisebox{0pt}{\tikz{\draw[-, color2, solid, line width = 1.0pt](0,0) -- (5mm,0);
\draw[color2,fill,line width = 1.0pt] (2.5mm, 0) circle (2pt);}}}
\newcommand{\Vmodelc}{\raisebox{0pt}{\tikz{\draw[-, color3, solid, line width = 1.0pt](0,0) -- (5mm,0);
\draw[color3,fill,line width = 1.0pt] (2.5mm, 0) circle (2pt);}}}
\newcommand{\Vmodeld}{\raisebox{0pt}{\tikz{\draw[-, color4, solid, line width = 1.0pt](0,0) -- (5mm,0);
\draw[color4,fill,line width = 1.0pt] (2.5mm, 0) circle (2pt);}}}
\newcommand{\Vmodele}{\raisebox{0pt}{\tikz{\draw[-, color5, solid, line width = 1.0pt](0,0) -- (5mm,0);
\draw[color5,fill,line width = 1.0pt] (2.5mm, 0) circle (2pt);}}}
\newcommand{\Vmodelf}{\raisebox{0pt}{\tikz{\draw[-, color6, solid, line width = 1.0pt](0,0) -- (5mm,0);
\draw[color6,fill,line width = 1.0pt] (2.5mm, 0) circle (2pt);}}}
\newcommand{\Vmodelg}{\raisebox{0pt}{\tikz{\draw[-, color7, solid, line width = 1.0pt](0,0) -- (5mm,0);
\draw[color7,fill,line width = 1.0pt] (2.5mm, 0) circle (2pt);}}}

\newcommand{\Liuetala}{\raisebox{0pt}{\tikz{\draw[blue,fill = blue] (2.5mm, 0) circle (1.75pt);}}}
\newcommand{\AP}{\raisebox{0pt}{\tikz{\draw[color4,fill = color4,scale=0.35] (0, 0) -- (0.5,0) --(0.25,-0.433) -- (0,0);}}}
\newcommand{\LS}{\raisebox{0pt}{\tikz{\draw[color5,fill = color5,scale=0.35] (0, 0) -- (0.5,0) --(0.5,-0.5) -- (0,-0.5);}}}
\newcommand{\SBLLESGO}{\raisebox{0pt}{\tikz{\draw[red,fill = red,scale=0.35,rotate=-90] (0, 0) -- (0.5,0) --(0.25,-0.433);}}}
\newcommand{\CNBLLESGO}{\raisebox{0pt}{\tikz{\draw[MATLAB_green,fill = MATLAB_green,scale=0.35,rotate=90] (0, 0) -- (0.5,0) --(0.25,-0.433);}}}

\newcommand{\redline}{\raisebox{2pt}{\tikz{\draw[-,red,solid,line width = 1.0pt](0,0) -- (5mm,0);}}}
\newcommand{\blueline}{\raisebox{2pt}{\tikz{\draw[-,blue,solid,line width = 1.0pt](0,0) -- (5mm,0);}}}
\newcommand{\bluedashedline}{\raisebox{2pt}{\tikz{\draw[-,blue,dotted,line width = 1.5pt](0,0) -- (5mm,0);}}}

\hypersetup{
	colorlinks=true,
	linkcolor=red,
	filecolor=magenta,    
	citecolor=blue,  
	urlcolor=cyan,
}
\newcommand{\myappendix}{
  \appendix
  \renewcommand{\thesection}{\appendixname~\Alph{section}}
}

\allowdisplaybreaks
\usepackage{multirow}

\newcommand*\patchAmsMathEnvironmentForLineno[1]{%
\expandafter\let\csname old#1\expandafter\endcsname\csname #1\endcsname
\expandafter\let\csname oldend#1\expandafter\endcsname\csname end#1\endcsname
\renewenvironment{#1}%
{\linenomath\csname old#1\endcsname}%
{\csname oldend#1\endcsname\endlinenomath}}%
\newcommand*\patchBothAmsMathEnvironmentsForLineno[1]{%
\patchAmsMathEnvironmentForLineno{#1}%
\patchAmsMathEnvironmentForLineno{#1*}}%
\AtBeginDocument{%
\patchBothAmsMathEnvironmentsForLineno{equation}%
\patchBothAmsMathEnvironmentsForLineno{align}%
\patchBothAmsMathEnvironmentsForLineno{flalign}%
\patchBothAmsMathEnvironmentsForLineno{alignat}%
\patchBothAmsMathEnvironmentsForLineno{gather}%
\patchBothAmsMathEnvironmentsForLineno{multline}%
}
\titlerunning{Analytical coupled Ekman and surface layer model for ABL flows}

\begin{document}

\title{Analytical model coupling Ekman and surface layer structure  in  atmospheric boundary layer flows}

\author{Ghanesh Narasimhan         \and
        Dennice F. Gayme \and Charles Meneveau
}

\institute{Ghanesh Narasimhan \at
              ghanesh91@gmail.com        
           \and
           Dennice F. Gayme \at
              dennice@jhu.edu
            \and
            Charles Meneveau \at
            meneveau@jhu.edu
}

\date{Received: DD Month YEAR / Accepted: DD Month YEAR}

\maketitle

\begin{abstract}
We  introduce  an analytical model that describes the vertical structure of Ekman boundary layer flows coupled to the MOST surface layer representation, which is valid for conventionally neutral (CNBL) and stable (SBL) atmospheric conditions. The model is based on a self-similar total stress distribution for both CNBL and SBL flows that merges the classic 3/2 power law profile  with a MOST-consistent stress profile in the surface layer. The velocity profiles are then obtained from the Ekman momentum balance equation. The same stress model is used to derive a new self-consistent Geostrophic Drag Law (GDL). We determine the ABL depth $h$ using an equilibrium boundary layer height model and parameterize the surface heat flux for quasi-steady SBL flows as a function of a prescribed surface temperature cooling rate. The ABL height and GDL equations can then be solved together to obtain the friction velocity $(u_*)$ and the cross-isobaric angle ($\alpha_0$) as a function of known input parameters such as the Geostrophic wind velocity magnitude and surface roughness $z_0$.  We show that the model predictions agree well with predictions from the literature and newly generated  Large Eddy Simulation (LES). These results indicate that the proposed model provides an efficient and reasonably accurate self-consistent approach for predicting the mean wind velocity distribution in CNBL and SBL flows.

\end{abstract}

\section{Introduction}
\label{intro}

The atmospheric boundary layer (ABL) refers to the lower part of the atmosphere, which interacts with processes on the Earth's surface, involving exchanges of heat, moisture, and momentum. 
In this region, Coriolis and frictional forces influence the vertical distribution of the wind. The resulting wind veers across different heights leading to an Ekman spiral flow \citep{ekman1905}. Understanding this wind veering, or Ekman spiral flow, under realistic atmospheric conditions is essential for applications such as wind energy, pollutant transport modeling, and boundary layer parameterizations for climate models. An example of this importance is highlighted in a study by \cite{walcek_2002}, which demonstrated that the presence of an Ekman spiral flow leads to a skewed structure in pollution puffs, challenging the traditional assumption of a simple shape for pollutant plumes.  
Wind veering in the ABL has similarly been shown to change the shape of wind turbine wake regions to be more sheared in the lateral direction, see e.g., \cite{Abkar_et_al_2018} and \cite{Narasimhan_et_al_2022}, versus the symmetric distribution assumed when neglecting the effect of  veer. This lateral extension of the wake due to wind veer can significantly impact the performance of downstream turbines. 
These and other findings underscore the importance of accounting for properties of ABL that influence wind farm flow physics and pollutant transport.

A number of studies have addressed the classical problem of modeling wind velocity distributions in the ABL. These velocities are typically obtained by solving the steady-state Ekman mean momentum equations. The traditional approach to solving these equations involves invoking the Boussinesq eddy-viscosity hypothesis to model turbulent shear stresses. \cite{ekman1905} solved these equations assuming a constant eddy-viscosity and obtained the classical solution involving a spiraling flow velocity profile based on trigonometric and exponential functions. However, in the turbulent ABL, the eddy-viscosity actually varies with height. In particular, the atmospheric surface layer (ASL), located close to the surface, exhibits a linear dependence of eddy-viscosity with height, assuming that the mixing length varies linearly with distance to the ground  \citep{Tennekes_Lumley_1972}. This layer, in which the constant turbulent shear stress is proportional to the square of the friction velocity $(u_*)$, is often referred to as the constant stress layer. Within the ASL, wind veering is negligible, and the streamwise velocity follows a logarithmic profile (if the effects of thermal stratification are neglected). 

Above the ASL lies the Ekman layer, which has flow scales that are large enough to be influenced by Coriolis effects leading to wind veering. Here, the logarithmic velocity profile that is valid in the ASL does not satisfy the Geostrophic wind condition approaching the top boundary. \cite{Ellison_1955} addressed this issue by solving the Ekman momentum equations with a linearly varying eddy-viscosity throughout the layer and obtained analytical solutions for velocities involving Kelvin and generalized Bessel functions \citep{krishna_ellison_1980}. These analytical solutions also showed that the vertical height $z$ is scaled by the thickness of the Ekman layer $h_e$, i.e. solutions depend on $z/h_e$, where $h_e=u_*/f_c$  \citep{Rossby_Montgomery_1935} and $f_c$ is the Coriolis frequency. However, the assumption of linearly increasing eddy-viscosity in \cite{Ellison_1955} is inaccurate within the Ekman layer. Instead the eddy-viscosity decays as it approaches the Geostrophic region, where turbulent stresses are negligible. \cite{Blackadar_1962} addressed this problem by
invoking Prandtl's mixing length theory to propose an eddy-viscosity profile that increases linearly within the ASL and decays within the Ekman layer. \cite{Blackadar_1962} solved the associated equations numerically to obtain ABL velocities that exhibit a logarithmic profile within the ASL and form an Ekman spiral structure further away from the surface, eventually merging with the Geostrophic wind. However, the form of these equations did not allow for analytical solutions. 

In a recent study, \cite{constantin_johnson_2019} showed that Ekman's mean momentum equations predict a spiraling velocity profile for any assumed form of eddy-viscosity that is bounded and reaches a constant value at larger heights. \cite{constantin_johnson_2019} explicitly showed that assuming a linearly varying eddy-viscosity closer to the surface and a constant value at greater heights leads to Bessel-type solutions for the mean velocity components. Similarly, assuming exponential eddy-viscosity variation closer to the surface that tends to a constant value away from it results in solutions given by hypergeometric functions. Although these explicit analytical solutions are instructive, evaluating such special functions becomes cumbersome and the complexity is often similar to having to  solve the eddy-viscosity momentum (1D ordinary differential) equations numerically. 

The effects of ground temperature add to the challenges of analytically describing the velocity profile in the ABL. Heating of the ground leads to the formation of an unstable or Convective Atmospheric Boundary Layer (CBL), while cooling the ground results in a Stable Atmospheric Boundary Layer (SBL). The Geostrophic free-stream region is stably stratified and separated from the ABL by a capping inversion layer, which typically forms at a height on the order of 1 km \citep{stull_1988}. When a neutral ABL exists beneath the stably stratified inversion layer and the free-stream flow, a type of boundary layer called the Conventionally Neutral Boundary Layer (CNBL) is formed. The strength of the wind veer in the Ekman layer depends on the atmospheric thermal stability. In particular, the wind veer is most pronounced in an SBL, while a more substantial vertical mixing resulting from convection in the CBL leads to weaker wind veering \citep{Deardorff_1972, wyngaard_2010,berg_et_al_2013,Liu_Stevens_RE_2022}. 

Within the lower part of the ABL, the ASL, analytical expressions for the velocity profiles including the effects of thermal stratification, can be obtained using Monin-Obukhov Similarity Theory (MOST) \citep{Monin_Obukhov_1954,dyer_1974}. The  MOST incorporates stability correction terms to account for the deviation from the logarithmic law behavior within the ASL due to the heating or cooling of the surface. Although the boundary layer region is neutrally stratified in CNBL flows, the stratification in the Geostrophic free-stream region influences the velocity profile in the surface layer. Studies such as \cite{Z_et_al_2002}, \cite{Taylor_Sarkar_2008}, and \cite{Abkar_Porteagel_2013} provide additional correction terms to the log law profile that incorporate the effect of free-stream stratification. However, these corrections
fail to accurately represent the flow within the Ekman outer layer, where wind veering is significant, and they do not typically satisfy the Geostrophic velocity condition at the ABL height.   \cite{gryning_et_al_2007} used a mixing length approach to obtain the mean velocity distribution above the surface layer in  cases including stratification.  
An additional layer was introduced where the mixing length decayed to zero around the ABL depth. Moreover the friction velocity was assumed to decay linearly with height within the ABL. Although the velocity profiles were constructed to match the Geostrophic wind, the emphasis was on prediction of velocity magnitude within the first few hundred meters of the ABL and the model did not provide analytical predictions of the separate velocity components. In \cite{kelly_gryning_2010}, the MOST and the Gryning model \citep{gryning_et_al_2007} were modified to include information about the probability distributions of the Monin-Obukhov length.

\cite{kadantsev_et_al_2021} reformulated the Ekman mean momentum equations in terms of the turbulent stresses and derived analytical expressions for the stresses. The analytical velocity profiles were then obtained by integrating the model stresses divided by an eddy-viscosity, which was modeled as the product of the friction velocity and a mixing length scale. Within the ASL, the mixing length was assumed to increase linearly with height until reaching a constant value in the Ekman layer. The constant value was determined based on an LES-tuned inverse quadratic expression for the turbulent length, which accounted for different atmospheric conditions. However, this assumption of a constant eddy-viscosity within the Ekman layer leads to inaccurate velocity predictions, particularly in the region approaching the top of the ABL. 

In addition to modeling the shape  of the velocity profile as a function of height,   means to predict the  friction velocity as a function of known Geostrophic velocity and surface roughness are also required. This is accomplished using what is often termed a Geostrophic Drag Law (GDL). \cite{kadantsev_et_al_2021} utilized LES to develop a  GDL model capable of predicting Geostrophic wind and friction velocities across a range of stability conditions, including CNBL and SBL flows.  In a recent study, \cite{Liu_Stevens_2022_PNAS} proposed  analytical expressions to predict both the streamwise and spanwise mean velocity components throughout the boundary layer height in CNBL flows. This study utilized a separate LES-based GDL model \citep{Liu_et_al_2021} to obtain the magnitudes of the Geostrophic wind and friction velocity for different CNBL flow conditions that are not directly based on the modeled velocity profile. Separately from their proposed GDL model, \cite{Liu_Stevens_2022_PNAS} provided stability correction functions  to predict the structure of the velocity components in the CNBL flows. Although its prediction of the full profile improves upon previous models that predicted only the resultant magnitude of the CNBL wind~\cite{kelly_et_al_2019}, \cite{Liu_et_al_2021_PRL}, the  \cite{Liu_Stevens_2022_PNAS} model was tailored to CNBL conditions. The extension of its applicability to SBL flows requires further development. 

Motivated by the limitations of existing models, specifically the need for more efficient and accurate velocity predictions in  CNBL and SBL flows, and a GDL that is self-consistently derived from the model of the velocity profile,  we propose a two-layer approach. In an outer (Ekman) layer,  we assume a self-similarity of turbulent stresses across stability conditions and a \cite{Nieuwstadt_1984} 3/2 power law profile. The second layer is an inner layer for the ASL that is consistent with the MOST description. Based on this description, we develop a self-consistent GDL to estimate the friction velocity and cross-isobaric angle by requiring continuity of mean velocities between the two layers. By incorporating the GDL into the model, we obtain the complete velocity profiles within the ABL for a variety of stability conditions (CNBL and a range of SBL). We validate our GDL and ABL wind model using data from the literature as well as from newly generated large-eddy simulations (LES).  The proposed analytical velocity distribution and GDL model are introduced in section \S\ref{Sec:2}. The LES datasets for model validation are described in section \S\ref{Sec:LES}, while results and discussions are presented in section \S\ref{Sec:Results}. Finally,  overall conclusions are summarized in section \S\ref{Sec:conclusion}.

\section{Analytical model for ABL wind velocity profiles}\label{Sec:2}

We employ a Cartesian coordinate system, with the $x$, $y$, and $z$ axes aligned with the streamwise (near the ground), spanwise, and wall-normal directions, respectively. Assuming a steady and horizontally homogeneous flow, we construct the model based on 
the streamwise and spanwise mean Ekman momentum equations:

\begin{align}
0 &= \mspace{15mu}f_c[V(z) - V_g] + \frac{\partial \mathrm{T}_{xz}(z)}{\partial z}, \label{mom_eq_x_dim}\\
0 &= -f_c[U(z) - U_g] + \frac{\partial \mathrm{T}_{yz}(z)}{\partial z}. \label{mom_eq_y_dim}
\end{align}
In these equations, $f_c$ represents the Coriolis frequency, while $U(z)$ and $V(z)$ correspond to the streamwise and spanwise components of the mean velocity, respectively. Additionally, $\mathrm{T}_{xz}(z)$ and $\mathrm{T}_{yz}(z)$ denote the turbulent stresses in the streamwise and spanwise directions, respectively. 
For simplicity, we derive the model assuming $f_c>0$, which corresponds to the northern hemisphere. A simple change in the sign of the spanwise components of the velocity and turbulent stress can be employed to model velocity profiles in the southern hemisphere.

The bottom boundary of the ABL is characterized by a surface roughness height $z_0$, where the velocities adhere to the no-slip velocity condition, $U(z_0) = V(z_0) = 0$. In the region near the surface as $z$ approaches $z_0$, a constant stress zone exists, where the turbulent stresses are predominantly aligned in the streamwise direction at the surface. Since the streamwise  direction near the surface is defined here as the $x$ direction, $\mathrm{T}_{xz}(z_0)=u_*^2$, where $u_*$ is the friction velocity. Since  $\mathrm{T}_{yz}(z_0) = 0$, and the resultant total surface stress is $\mathrm{T}(z_0)=\sqrt{\mathrm{T}_{xz}^2(z_0)+\mathrm{T}_{yz}^2(z_0)}=\mathrm{T}_{xz}(z_0)=u_*^2$ (It is straightforward to recast any results in an arbitrarily chosen coordinate system, as long as the Geostrophic wind direction is adjusted accordingly).  

The ABL  extends to a height of $h$.  Above $h$, the flow transitions into a region under Geostrophic balance, characterized by stable stratification and the absence of turbulent stresses. From \eqref{mom_eq_x_dim} and \eqref{mom_eq_y_dim}, the velocity components in this Geostrophic region can be simplified to $V(z \geq h) = V_g$ and $U(z \geq h) = U_g$. Assuming a Geostrophic wind with a magnitude of $G$ oriented at an angle $\alpha_0$ relative to the streamwise ($x$) direction, the velocity components can be expressed as $U_g = G\cos\alpha_0$ and $V_g = G\sin\alpha_0$.

The dimensionless form of equations \eqref{mom_eq_x_dim} and \eqref{mom_eq_y_dim}  utilize the friction velocity $u_*$ and the Rossby-Montgomery length scale \citep{Rossby_Montgomery_1935} $u_*/f_c$ as the characteristic velocity and length scales of the flow. 
The model to be developed in this paper is based on the dimensionless form of the mean momentum equations \eqref{mom_eq_x_dim} and \eqref{mom_eq_y_dim}. For known values of $U_g$, $V_g$, $u_*$, and given the profiles of the non-dimensional form of turbulent stresses $\T{T}_{xz}(\hat{\xi})=\mathrm{T}_{xz}(z)/u_*^2$, and $\T{T}_{yz}(\hat{\xi})=\mathrm{T}_{yz}(z)/u_*^2$, we can obtain the ABL wind velocity profiles as:

\begin{align}
\frac{V(z)}{u_*} &= -\frac{\partial \T{T}_{xz}(\hat{\xi})}{\partial \hat{\xi}} + \frac{V_g}{u_*}, \label{V_eqn} \\
\frac{U(z)}{u_*} &= \mspace{15mu}\frac{\partial \T{T}_{yz}(\hat{\xi})}{\partial \hat{\xi}} + \frac{U_g}{u_*}, \label{U_eqn}
\end{align}
where $\hat{\xi} = z f_c/u_*$ represents the dimensionless vertical coordinate. Correspondingly, $\hat{h}=h f_c/u_*$ represents the dimensionless ABL height and $\hat{\xi}_0=z_0f_c/u_*$ defines the dimensionless surface roughness length. 

In the following sections, we will describe the proposed model for the turbulent stresses ($\T{T}_{xz}$ and $\T{T}_{yz}$). This model will be used to self-consistently derive a new GDL to determine $U_g$, $V_g$, and $u_*$ from  the resulting velocity profiles. A classic form of the GDL  to obtain the friction velocity and cross-isobaric angle for given Geostrophic velocity $G$ and roughness length $z_0$ can be written as \citep{Rossby_Montgomery_1935,Tennekes_Lumley_1972,Zilitinkevich_Esau_2005,Liu_et_al_2021}

\begin{align}
    \frac{\kappa G\cos\alpha_0}{u_*}&=\ln\left(Ro\right)-A,\label{GDL_A}\\
    \frac{\kappa G\sin\alpha_0}{u_*}&=- B,\label{GDL_B}
\end{align}
where $Ro=u_*/(f_c z_0)$ is the friction Rossby number, $\kappa$ is the   Von-K\'arm\'an constant (often taken to be $\kappa \approx 0.41$), and $A,B$ are dimensionless parameters. Note that the resistive laws in equations \eqref{GDL_A},\eqref{GDL_B} are written here for the northern hemisphere, where $V_g<0$. For the southern hemisphere,  $V_g>0$ which is prescribed by having a positive sign before the coefficient $B$ in \eqref{GDL_B}. 

Previous studies have shown that the constants $A$ and  $B$ depend on the atmosphere's stability conditions \citep{Zilitinkevich_Esau_2005,kadantsev_et_al_2021}. In our current study, we develop new functional forms for the dimensionless constants $A$ and $B$ that can be used to self-consistently determine $u_*$ and $\alpha_0$ (and thus $U_g$ and $V_g$)  across neutral and stable atmospheric conditions.

\subsection{\textbf{Assumed turbulent stress distributions}}\label{Sec:stress_model}

\begin{figure}[t]
    \centering
\includegraphics{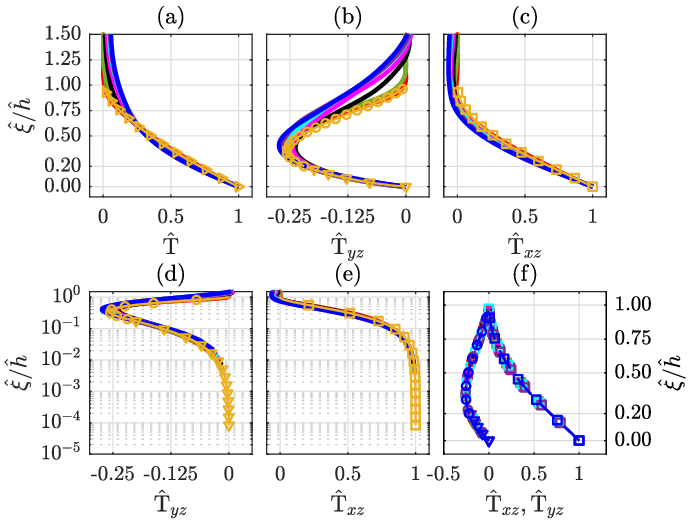}
    \caption{Profiles of normalized  turbulent stresses (a) $\T{T}$, (b)\&(d) $\T{T}_{yz}$, (c)\&(e) $\T{T}_{xz}$  from the LES (Table \ref{tab:LESGO}) of CNBL ($\protect\CNBL$), SBL-1 ($\protect\SBLa$), SBL-2 ($\protect\SBLb$), SBL-3 ($\protect\SBLc$), SBL-4 ($\protect\SBLd$), SBL-5 ($\protect\SBLe$), SBL-6 ($\protect\SBLf$) compared with analytical profiles of  $\T{T}$ ($\protect\Ttotmodel$, Eq. \ref{T_tot}), $\To{T}_{yz}$   ($\protect\model$, Eq. \ref{Tyz_outer}), $\Ti{T}_{yz}$ ($\protect\Tiyzmodel$, Eq. \ref{Tyz_inner}), $\T{T}_{xz}$ ($\protect\Txzmodel$, Eq. \ref{Txz_model}) generated using the parameters of the CNBL case in Table \ref{tab:LESGO}. 
    Plot (f) shows analytical profiles of 
    $\T{T}_{xz}$ 
    ($\protect\Txzmodelfa$,
    $\protect\Txzmodelfb$,
    $\protect\Txzmodelfc$,
    $\protect\Txzmodelfd$,
    $\protect\Txzmodelfe$,
    $\protect\Txzmodelff$,
    $\protect\Txzmodelfg$),
    $\T{T}_{yz}$ 
    ($\Ti{T}_{yz}$ : 
    $\protect\Tiyzmodelfa$, 
    $\protect\Tiyzmodelfb$,
    $\protect\Tiyzmodelfc$,
    $\protect\Tiyzmodelfd$,
    $\protect\Tiyzmodelfe$,
    $\protect\Tiyzmodelff$,
    $\protect\Tiyzmodelfg$,
    $\To{T}_{yz}$ : 
    $\protect\Toyzmodelfa$,
    $\protect\Toyzmodelfb$,
    $\protect\Toyzmodelfc$,
    $\protect\Toyzmodelfd$,
    $\protect\Toyzmodelfe$,
    $\protect\Toyzmodelff$,
    $\protect\Toyzmodelfg$) generated using the LES parameters in Table \ref{tab:LESGO} following the same color code as the LES plots in panels (a)-(e).
    } 
    \label{stress_plot}
\end{figure}

Previous studies \cite{Nieuwstadt_1984}, \cite{Zilitinkevich_Esau_2005}, \cite{Liu_et_al_2021_PRL} have investigated the vertical profiles of the total stress $\T{T}$ within the CNBL and SBL flows. They propose the following analytical representation for the total stress $\T{T}=\sqrt{\T{T}_{xz}^2+\T{T}_{yz}^2}$:

    \begin{align}
    \T{T}=\left(1- {\hat{\xi}}/{\hat{h}}\right)^{3/2}. \label{T_tot}
    \end{align}
Figure \ref{stress_plot}(a) confirms good agreement of this proposed self-similar profile with data from Large Eddy Simulations (LES) of CNBL and SBL flows  (details about the LES are provided in   \S\ref{Sec:LES}).  Here results for various stability conditions are plotted as solid lines in normalized form and the analytical \cite{Nieuwstadt_1984} model for total stress is denoted  using yellow right-pointing triangle markers on a solid yellow line. The dimensional estimate of the ABL depth $h$ is determined using a least-square error fitting of the 3/2-power law curve in equation \eqref{T_tot} with the total turbulent stress from the LES. The dimensional ABL depth for the LES cases plotted in Figure \ref{stress_plot} are given in Table \ref{tab:LESGO}.  Figures \ref{stress_plot}(b,c) show results for the individual stress components, also showing reasonably good self-similar collapse in the ABL region when using the fitted boundary layer height $\hat{h}$ to normalize the $\hat{\xi}$ axis.  In order to develop an analytical expression for the velocity profiles from Eqs. \eqref{V_eqn} \& \eqref{U_eqn}, an additional model is needed for the  spanwise turbulent stress component $\hat{\mathrm{T}}_{yz}$. The streamwise one can then be obtained by ensuring the total stress is given by Eq. \eqref{T_tot}.

As discussed in the introduction, two distinct layers can be identified in the ABL: the outer or Ekman layer, and the inner or surface (ASL) layer. The outer layer is characterized by Coriolis effects and the gradual decay of turbulent stresses to zero as the ABL height $h$ is approached. Motivated by the LES observations in 
Figure \ref{stress_plot}(b) we propose the following  model of the spanwise turbulent stress in this outer region

    \begin{align}
    \To{T}_{yz} = - \ g(\hat{\xi})\left(1- {\hat{\xi}}/{\hat{h}}\right)^{3/2}, \label{Tyz_outer} \,\,\,\, {\rm where} \,\,\,\, 
    g(\hat{\xi}) = M\left(1-e^{-\Gamma\hat{\xi}/\hat{h}}\right). 
    \end{align}
    Here, $M=1.43$ is a fitting parameter obtained by calibrating the model with the results of the LES, see Figure \ref{stress_plot}(b). The constant $M$ determines the magnitude of the turbulent spanwise stress and $\Gamma=1.2$ is chosen to reproduce the decay and diminishing contribution of the outer layer  as $\hat{\xi}$ approaches $\hat{\xi}_0$ also ensuring that $g(\hat{\xi})<1$ as $\hat{\xi}$ approaches $\hat{h}$. The outer layer is assumed to be valid for $\hat{\xi}>\hat{\xi}_m$, i.e. above a certain transition height $\hat{\xi}_m$ to be specified later but that is typically expected to be 10-20\% of the boundary layer height $\hat{h}$.
        
    For $\hat{\xi} \leq \hat{\xi}_m$, the inner layer of the ABL consists of the log layer and the $z$-independent stratification layer \citep{Zilitinkevich_Esau_2005}. Within this region, the wind veer flow is negligible, and the streamwise velocity for  SBL (and CNBL) flows is governed by the Monin-Obukhov Similarity Theory (MOST) \citep{Monin_Obukhov_1954}, which can be expressed as:
    
    \begin{align}
        \frac{U^{\text{inner}}}{u_*}&=\frac{1}{\kappa}\left(\ln\frac{z}{z_0}+5\,\frac{z-z_0}{L_s}\right). \label{U_MOST}
    \end{align}
    
    Here $L_s$ is the Monin-Obukhov length scale defined as
    
    \begin{align}
    L_s &= -\frac{u_*^3}{\kappa (g/\Theta_0)Q_0}, \label{Obukhov_len}
    \end{align}
    where $Q_0$ denotes the surface cooling flux and $\Theta_0$ represents the reference potential temperature scale. Eq. \eqref{U_MOST} indicates that for the SBL flows, the velocity profile approximates the log layer very close to the ground since the log term becomes dominant as $z$ approaches $z_0$. Similarly, as we move further away from $z_0$, the stability correction term becomes dominant and the ABL velocity follows the $z$-independent stratification behavior in which the velocity gradient is independent of $z$ ($\partial U^{\text{inner}}/\partial z\approx 5/L_s$). The pure logarithmic law for the CNBL conditions is recovered for $|L_s| \to \infty$. Using the dimensionless wall-normal distance $\hat{\xi}=zf_c/u_*$, equation \eqref{U_MOST} can be expressed as
    \begin{equation}
    \frac{U^{\text{inner}}}{u_*} = \frac{1}{\kappa} \ln\frac{\hat{\xi}}{\hat{\xi}_0} + 5\mu (\hat{\xi}-\hat{\xi}_0), \label{U_inner_non_dim_SBL}
    \end{equation}
    where $\mu$ is the stability parameter defined as
    
    \begin{align}
    \mu &= \frac{u_*}{\kappa f_c L_s}.\label{mu_param}
    \end{align}
        
    \cite{Abkar_Porteagel_2013} performed an LES study that investigated the effect of free-stream stratification on the velocity structure in the surface layer of the CNBL. That study suggested that the velocity profile can be expressed as:
    
    \begin{align}
        \frac{U^{\text{inner}}}{u_*}&=\frac{1}{\kappa}\ln\frac{\hat{\xi}}{\hat{\xi}_0}+0.3 \mu_N (\hat{\xi}-\hat{\xi}_0), \label{U_inner_non_dim_CNBL}
    \end{align}
   where, as before $\hat{\xi}=zf_c/u_*$ and
     $\mu_N$ is the Zilitinkevich number defined as
    \begin{align}
        \mu_N=\frac{N_\infty}{f_c}.  \label{muN_param}
    \end{align}
    Here, $N_\infty=\sqrt{(g/\Theta_0) \gamma_\Theta}$ is the Brunt-V\"ais\"al\"a frequency associated with the potential temperature that varies with a lapse rate of $\gamma_\Theta$ (K/m) in the free stream region.
    
    In order to represent  $U^{\text{inner}}$ for  both the SBL and CNBL flows, we combine equations \eqref{U_inner_non_dim_SBL}, \eqref{U_inner_non_dim_CNBL} and express $U^{\text{inner}}$ as:
    
    \begin{align}
        \frac{U^{\text{inner}}}{u_*}&=\frac{1}{\kappa}\ln\frac{\hat{\xi}}{\hat{\xi}_0}+(5\mu+0.3\mu_N)(\hat{\xi}-\hat{\xi}_0). \label{U_inner}
    \end{align}
    
    To determine associated inner layer stress, we substitute equation \eqref{U_inner} into the $y$-momentum equation \eqref{U_eqn} and integrate along $\hat{\xi}$ to obtain  the transverse stress profile 
    
    \begin{align*}
        \Ti{T}_{yz}&=\dfrac{\hat{\xi}}{\kappa} \ln \dfrac{\hat{\xi}}{\hat{\xi}_0}-\dfrac{1}{\kappa}(\hat{\xi}-\hat{\xi}_0)\Biggl[1+\kappa\dfrac{U_g}{u_*}-\dfrac{\kappa}{2}(5\mu+0.3 \mu_N)(\hat{\xi}-\hat{\xi}_0)\Biggr].\numberthis\label{Tyz_inner}
    \end{align*}
    Then combining equations \eqref{Tyz_outer} and  \eqref{Tyz_inner}, the spanwise turbulent stress $\T{T}_{yz}$ is modeled as:
    
    \begin{align*}
    \hat{\mathrm{T}}_{yz}&=
    \begin{cases}
    - \ g(\hat{\xi})\left(1-\dfrac{\hat{\xi}}{\hat{h}}\right)^{3/2} & \hat{\xi}>\hat{\xi}_m\\
    &\\
    \dfrac{\hat{\xi}}{\kappa} \ln \dfrac{\hat{\xi}}{\hat{\xi}_0}-(\hat{\xi}-\hat{\xi}_0)\Biggl[\dfrac{1}{\kappa}+\dfrac{U_g}{u_*}
    -\dfrac{1}{2}(5\mu+0.3 \mu_N)(\hat{\xi}-\hat{\xi}_0)\Biggr] &  \hat{\xi}<\hat{\xi}_m
    \end{cases}.\numberthis \label{Tyz_model}
    \end{align*}
    
    Note that in this equation, the value of $U_g$ is not yet determined and will be found self-consistently later by enforcing continuity of velocities at the matching height $\hat{\xi}_m$. 
    Based on LES results and consistent with usual expectations that the inner solution is valid in the lower 10-20\% of the boundary layer height, we find that setting $\hat{\xi}_m = c_m \hat{h}$ with $c_m=0.20$ leads to good results. 
    The outer layer stress $\To{T}_{yz}$ from equation \eqref{Tyz_outer} and the inner layer stress $\Ti{T}_{yz}$ from equation \eqref{Tyz_inner} are plotted in Figure \ref{stress_plot}(b) using yellow circle and downward-pointing triangle markers on a solid yellow line, respectively. Both plots demonstrate good agreement with the LES data. Figures \ref{stress_plot}(d,e) use semilogarithmic axes to highlight the behavior of the inner portion near the ground.
    
    In these comparisons, we have used the value of $U_g/u_*$ determined from the proposed Geostrophic drag law (Eq. \ref{GDL_A} with Eq. \ref{A_model_GDL} to be developed in section \S \ref{Sec:GDL_new}). The model is evaluated for the same $z_0=0.1$ m as the LES and the cases shown correspond to $\mu =[0,5.62,20.59,39.84,59.25,78.35,148.49]$ and $\mu_N=61$ (see Table \ref{tab:LESGO}).
    Note that the analytical stress profiles in Figure \ref{stress_plot}(a)-(e) are plotted using the CNBL LES parameters, $\mu=0, \ \mu_N=61$. Figure \ref{stress_plot}(f) shows the analytical profiles of the turbulent stresses generated using all the LES parameters covering both the CNBL and SBL flows. We observe that these analytical turbulent stress components from the various LES cases lie on top of each other in Figure \ref{stress_plot}(f) with negligible differences occurring near the matching height $\hat{\xi}_m$.
    More detailed comparisons between the model and LES will be shown later in terms of velocity distributions.

The distribution of the normalized streamwise turbulent stress $\T{T}_{xz}$ is obtained directly from the definition of the total stress  as follows:

    \begin{align}
        \T{T}_{xz}&=\sqrt{\T{T}^2-\T{T}_{yz}^2} \label{Txz}
    \end{align}
    We can then use the model expressions for $\T{T}$ from equation \eqref{T_tot},  and $\T{T}_{yz}$ from equation \eqref{Tyz_model} in \eqref{Txz} to obtain  $\T{T}_{xz}$. For this component, an inner portion is not required: From equation \eqref{Tyz_inner}, we observe that $\Ti{T}_{yz}(\hat{\xi})$ tends to zero as $\hat{\xi}$ approaches $\hat{\xi}_0$. This shows that $\Ti{T}_{yz}\ll \T{T}$ within the surface (inner) layer. In addition, since $\To{T}_{yz}(\hat{\xi}_0)=0$, we model $\T{T}_{xz}$   by using only $\To{T}_{yz}$ in equation \eqref{Txz}:

    \begin{align*}
    \T{T}_{xz}=\sqrt{\T{T}^2-(\To{T}_{yz})^2} \,=\,
    \sqrt{1-g(\hat{\xi})^2}\left(1-\frac{\hat{\xi}}{\hat{h}}\right)^{3/2}.\numberthis \label{Txz_model}
    \end{align*}

    This analytical profile of $\T{T}_{xz}$ is plotted as yellow square markers on a yellow solid line in Figures \ref{stress_plot}(c)\&(e), again showing good agreement with the LES.
    Using these turbulent stress models in the Ekman mean momentum equations, we obtain the ABL wind velocity profiles in the following section.

\subsection{\textbf{ABL velocity profiles}}

We obtain the analytical expressions for the ABL velocity profiles using the modeled stresses from \eqref{Txz_model} and \eqref{Tyz_model} in \eqref{V_eqn} and \eqref{U_eqn}, respectively and evaluating their derivatives analytically. The result is:  

\begin{align}
    \frac{U(z)}{u_*}&=
    \begin{cases}
    -g^\prime(\hat{\xi})\left(1-\dfrac{\hat{\xi}}{\hat{h}}\right)^{3/2}+g(\hat{\xi})\dfrac{3}{2\hat{h}}\left(1-\dfrac{\hat{\xi}}{\hat{h}}
    \right)^{1/2}+\dfrac{U_g}{u_*}
    &, \ \hat{\xi}\geq\hat{\xi}_m\\
    &\\
    \dfrac{1}{\kappa}\ln\dfrac{\hat{\xi}}{\hat{\xi}_0}+(5\mu+0.3\mu_N)(\hat{\xi}-\hat{\xi}_0) &,\ \hat{\xi}\leq\hat{\xi}_m
    \end{cases},\label{U_eqn_model}\\
    \frac{V(z)}{u_*}&=\dfrac{g(\hat{\xi})g^\prime(\hat{\xi})}{\sqrt{1-g(\hat{\xi})^2}}\left(1-\dfrac{\hat{\xi}}{\hat{h}}\right)^{3/2}+\dfrac{3}{2\hat{h}}\sqrt{1-g(\hat{\xi})^2}\left(1-\dfrac{\hat{\xi}}{\hat{h}}\right)^{1/2}+\dfrac{V_g}{u_*},\label{V_eqn_model}
\end{align}
where $g^\prime(\hat{\xi})$ is the  derivative of $g(\hat{\xi})$  

\begin{align}
    g^\prime(\hat{\xi}) & =\frac{\Gamma M}{\hat{h}}e^{-\Gamma \hat{\xi}/\hat{h}}.\label{dfdx_func}
\end{align}
The streamwise velocity in \eqref{U_eqn_model} can be applied to both the northern and southern hemispheres. 
For the spanwise velocity in the southern hemisphere, a sign change is required for the first two terms on the right-hand side of \eqref{V_eqn_model} . The sign of the Geostrophic velocity component $V_g$ is set by the GDL equation \eqref{GDL_B}.
(As an aside, we note that recently \cite{kelly_laan-2023} use the flow's shear exponent, i.e., a power-law rather than the logarithmic description inherent in the MOST approach, to derive analytical expressions linking directional shear to the flow's shear exponent and the vertical gradient of cross-wind stress components.) 

In order for the model equations \eqref{U_eqn_model} and \eqref{V_eqn_model} to be complete, we still need to specify the ABL height $\hat{h}$, friction velocity $u_*$, Geostrophic velocity components $U_g/u_*$ and $V_g/u_*$, and stability parameter $\mu$. The latter requires $L_s$, which needs specification of the surface cooling flux $Q_0$.

\subsection{\textbf{Model for surface cooling flux}}
 
In this section, we aim to relate $Q_0$ to the imposed cooling rate $C_r=\partial \Theta_s/\partial t$ of the surface potential temperature $(\Theta_s)$. The governing equation for the evolution of the mean potential temperature $(\Theta)$ for the SBL flows is given by

\begin{align}
    \frac{\partial \Theta}{\partial t}&=-\frac{\partial \mean{w^\prime\theta^\prime}}{\partial z},\label{mean_T_eqn}
\end{align}
where $\langle w^\prime \theta^\prime\rangle$ represents the turbulent heat flux.

When the SBL reaches a quasi-stationary state, the difference between the potential temperature within the boundary layer and the surface temperature, $\Theta(t)-\Theta_s(t)$ is steady \citep{Brost_Wyngaard_1978}. Using this quasi-steady assumption, equation \eqref{mean_T_eqn} can be written as

\begin{align*}
    \frac{\partial}{\partial t}(\Theta-\Theta_s)+\frac{\partial \Theta_s}{\partial t}&=-\frac{\partial \mean{w^\prime\theta^\prime}}{\partial z},
    \implies \frac{\partial \mean{w^\prime\theta^\prime}}{\partial z}=-C_r.\numberthis\label{heat_flux_gradient}
\end{align*}
Integrating equation \eqref{heat_flux_gradient} across the SBL depth, we get an analytical estimate for the surface cooling flux $Q_0$:

\begin{align*}
    \int_{z_0}^{h}\frac{\partial \mean{w^\prime\theta^\prime}}{\partial z} \ dz &=-\int_{z_0}^{h} C_r \, dz, \implies
    Q_0\equiv\mean{w^\prime \theta^\prime}_s=C_r h \numberthis \label{Q0_model}.
\end{align*}
In obtaining equation \eqref{Q0_model}, we have used $\langle w^\prime \theta^\prime\rangle_h=0$, the top boundary condition for the heat flux. Using \eqref{Q0_model} in \eqref{Obukhov_len}, the Monin-Obukhov lengthscale $L_s$ for the SBL flows can be modeled as
\begin{align*}
    L_s&=\frac{u_*^3 \Theta_0}{\kappa g \, (-C_r) h}.\numberthis\label{Ls_model}
\end{align*}
As a result, we model the stability parameter $\mu$ as follows:

\begin{align*}
    \mu&=\frac{u_*}{\kappa f_c L_s}=\frac{g \, (-C_r)}{u_* f_c^2 \Theta_0} \frac{h f_c}{u_*}=\frac{g \, (-C_r)}{u_* f_c^2 \, \Theta_0} \hat{h}. \numberthis\label{mu_model}
\end{align*}

In our current study, we perform the LES of SBL flows using cooling rates $C_r=[0,-0.03,-0.125,-0.25,-0.375,-0.5,-1]$ K/hr (see Table \ref{tab:LESGO}). The corresponding stability parameter $\mu$ ranges between 0 to 148.49. The comparison of the model estimates for $Q_0$ \eqref{Q0_model} and stability parameter $\mu$ \eqref{mu_model} with the LES values are discussed in detail in section \S\ref{Sec:GDL_comparison}. This expression to calculate $\mu$ is used in the evaluation of the entire velocity profile  $U(z)$  from equation \eqref{U_eqn_model} (it does not directly affect the profile $V(z)$  although indirectly it affects the profile via its dependence on $u_*$, which is affected by the value of $\mu$).

The model expressions for $\mu$ and velocity profiles depend on the non-dimensional ABL height $\hat{h}$. In the following section, we discuss the model expression for determining the ABL depth for a given atmospheric stability condition.

\subsection{\textbf{ABL height model}}

We utilize a well-established equilibrium ABL height model, which has been extensively studied in previous works such as \cite{Z_et_al_2007} and \cite{Liu_et_al_2021}, to determine the non-dimensional ABL height. This model is expressed as follows:

\begin{align}
\frac{1}{\hat{h}^2} = \frac{1}{C_{TN}^2} + \frac{\mu_N}{C_{CN}^2} + \frac{\mu}{C_{NS}^2},\label{h_hat_model}
\end{align}
where $\mu$ is the stability parameter and $\mu_N$ is the Zilitinkevich number given by equations \eqref{mu_param} and \eqref{muN_param}, respectively.

This expression represents a smooth merging of  dimensionless height models corresponding separately to the Truly Neutral Boundary Layer (TNBL), CNBL, and SBL flows. \cite{Liu_et_al_2021} performed a suite of LES of the CNBL flows and obtained 
$C_{TN}=0.5$ and 
$C_{CN}=1.6$ by fitting to LES data, and we adopt those values as well. Similarly, we determine the model parameter $C_{NS}=0.78$ via fitting results from the LES of the SBL flows performed as part of our current study.

Incorporating the model for $\mu$ from  Eq. \eqref{mu_model}, we can re-write the expression for $\hat{h}$ as

\begin{align}
    \frac{1}{\hat{h}^3}=\frac{1}{\hat{h}}\left[\frac{1}{C_{TN}^2}+\frac{\mu_N}{C_{CN}^2}\right]+\frac{1}{C_{NS}^2}\frac{g\,(-C_r)}{u_* f_c^2 \Theta_0}.\label{h_hat_mu_model}
\end{align}

Using Eq. \eqref{h_hat_mu_model} with these empirically determined model parameters along with specifying the parameters $\mu_N$ and $C_r$, we can compute the non-dimensional ABL height $\hat{h}$  spanning different conventionally neutral and stable atmospheric stability conditions. The   cubic equation \eqref{h_hat_mu_model} can be solved e.g., iteratively (more details provided below). However, it is evident from Eq. \eqref{h_hat_mu_model} that we also require knowledge of the friction velocity $u_*$. We estimate $u_*$ from the new Geostrophic drag law model discussed in the following section. 

\subsection{\textbf{New Geostrophic Drag Law}\label{Sec:GDL_new}}

Completing the analytical model requires a method to determine $u_*$ and the flow angle $\alpha_0$ for given flow conditions and surface roughness $z_0$. We follow the classical approach of matching the inner (MOST) solution to outer conditions. In our context, we match the streamwise velocity profiles in Eq. \eqref{U_eqn_model} at the matching height $\hat{\xi}_m=c_m \hat{h}$, which enables us to determine  $U_g/u_*$  as follows: 

\begin{align*}
        \frac{U_g}{u_*}&=\frac{1}{\kappa}\ln\frac{c_m\hat{h}}{\hat{\xi}_0}+(5\mu+0.3\mu_N)(c_m\hat{h}-\hat{\xi}_0)\\
        &\mspace{20mu}+g^\prime(\hat{\xi}_m)\left[1-c_m\right]^{3/2}-g(\hat{\xi}_m)\frac{3}{2\hat{h}}\sqrt{1-c_m}.\numberthis\label{Ug_hat}
\end{align*}
For $V(z)$ no inner layer was required and thus the matching can be done by evaluating the spanwise velocity at the surface roughness height $\hat{\xi}=\hat{\xi}_0$. The result is 

\begin{align*}
    \frac{V_g}{u_*}=-\frac{3}{2\hat{h}}.\numberthis\label{Vg_hat}
\end{align*}

Since $U_g=G\cos\alpha_0$ and $V_g=G\sin\alpha_0$, comparing equations \eqref{Ug_hat} and \eqref{Vg_hat} with the classical GDL expressions from Eqs. \eqref{GDL_A} and \eqref{GDL_B} for the northern hemisphere
we get the following new expressions for the coefficients $A$ and $B$: 

\begin{align*}
A&=-\ln c_m \hat{h}-\kappa\Biggl[(5\mu+0.3\mu_N)(c_m\hat{h}-\hat{\xi}_0)+g^\prime(\hat{\xi}_m)\left(1-c_m\right)^{3/2}\\
&\mspace{20mu}-g(\hat{\xi}_m)\frac{3}{2\hat{h}}\sqrt{1-c_m}\Biggr],\numberthis\label{A_model_GDL}\\
B&=\frac{3\kappa}{2\hat{h}}.\numberthis\label{B_model_GDL}
\end{align*}

The equations \eqref{GDL_A}, \eqref{GDL_B}, \eqref{A_model_GDL}, \eqref{B_model_GDL} together constitute the new Geostrophic drag law model. Note that the coefficients $A$ and $B$ depend on $\hat{h}$ and the stability parameters, $\mu_N$, and $\mu$. The presence of these stability parameters allows for the estimation of $u_*$ and $\alpha_0$ for both the CNBL and SBL flows. However, obtaining these estimates is challenging due to the non-linear and interdependent nature of the GDL equations when coupled with the model expressions for $\hat{h}$ Eq. \eqref{h_hat_mu_model} and $\mu=g\,(-C_r)\hat{h}/(u_* f_c^2 \Theta_0)$ Eq. (\ref{mu_model}). As a result, obtaining entirely closed-form analytical solutions is not possible. One must solve the set of equations numerically and we propose a simple iterative approach to obtain $h,u_*,\alpha_0$.

We implement this iterative solution using a two-step process. Starting from an initial guess for $h$ and $u_*$, we first solve for $h$ and $u_*$ from the ABL depth model Eq. (\ref{h_hat_mu_model}) and the GDL equations Eqs. \eqref{GDL_A}, \eqref{GDL_B}, \eqref{A_model_GDL}, \eqref{B_model_GDL}. Eliminating $\alpha_0$ from the GDL equations, one obtains:   

  \begin{align}
        u_*&=\frac{\kappa G}{\sqrt{[\ln (Ro)-A]^2+B^2}}.\label{u_tau_model}
    \end{align}
Here $A$ and $B$ are from Eqs. \eqref{A_model_GDL}, \eqref{B_model_GDL}. The model for $\mu$ from Eq. \eqref{mu_model} is also utilized by the expression for $A$. We obtain dimensional values of $h,u_*$ by iteratively solving equations \eqref{h_hat_mu_model} and \eqref{u_tau_model}. 
    
From the dimensional estimates of $h$, $u_*$ found in the previous step, we can evaluate  $\hat{h}=h f_c/u_*$, $\hat{\xi}_0=z_0 f_c/u_*$  and $\mu=g \, (-C_r)\hat{h}/(u_* f_c^2 \Theta_0)$. Using these results, we can obtain the GDL coefficients $A$ and $B$. Given these coefficients, we evaluate $U_g$ and $V_g$ from Eqs. \eqref{GDL_A} and \eqref{GDL_B}, respectively. Finally, from the Geostrophic velocities, we find the cross-isobaric angle using $\alpha_0=\tan^{-1}(V_g/U_g)$. This iterative process of evaluating $h,u_*,\alpha_0$ is explained in more detail in  \ref{Sec:appendix}.

\section{Description of Large Eddy Simulation data for model validation}\label{Sec:LES}

Large Eddy Simulations have played a major role in improving our understanding of the ABL under various stability conditions \citep{saiki2000large,beare2006intercomparison,kumar2006large}. 
This section describes the various LES datasets used to validate our analytical model. Prior LES-based research has investigated several parameterizations of resistive laws and wind profiles for a  CNBL flow, see e.g., \cite{Abkar_Porteagel_2013}, \cite{Liu_et_al_2021}, \cite{Liu_et_al_2021_PRL}, and \cite{Liu_Stevens_2022_PNAS}. We describe the CNBL data from this existing literature in section \S\ref{Sec:CNBL_lit}.  We perform additional LES runs including SBL flows in this study, which we describe in section \S\ref{Sec:SBL_data}.   We utilize the ABL depth ($h$), friction velocity ($u_*$), and cross-isobaric angle ($\alpha_0^\circ$) values obtained in the prior studies and these new LES to validate our GDL model.

\subsection{LES data of CNBL flows from existing literature\label{Sec:CNBL_lit}}
In these previous studies, the LES of the CNBL was set up with a linear initial potential temperature profile, given by $\Theta(z)=\Theta_0+\gamma_\Theta z$. Here, $\gamma_\Theta$ represents the lapse rate of potential temperature in the free-stream Geostrophic region, and $\Theta_0$ is a reference potential temperature. For the initial velocity profiles, \cite{Abkar_Porteagel_2013} employed a laminar flow with random perturbations added near the surface within the first 100 m to initiate turbulence. Similarly, \cite{Liu_et_al_2021}, \cite{Liu_et_al_2021_PRL}, and \cite{Liu_Stevens_2022_PNAS} utilized a uniform velocity profile with a Geostrophic wind magnitude $G$, along with random perturbations within the first 100 m from the surface. By imposing a thermally insulating boundary condition at the bottom wall, a quasi-steady CNBL flow is established, characterized by a mean potential temperature exhibiting a capping inversion layer that separates the neutral boundary layer region from the Geostrophic free-stream region.

In \cite{Abkar_Porteagel_2013}, the CNBL flow was examined with a Geostrophic wind magnitude of $G=10$ m/s. Two surface roughness heights, namely $z_0=0.01 \ \text{m}$ and $z_0=0.1 \ \text{m}$ were considered while the Coriolis frequency was fixed to $f_c=10^{-4}$ 1/s. Two free-stream stratification strengths were chosen to yield Zilitinkevich numbers of $\mu_N=58$ and $\mu_N=180$. A total of four LES cases were performed in that study. The specific values for the CNBL depth, friction velocity, and cross-isobaric angle corresponding to these cases are listed in Table \ref{tab:data_CNBL}. The study's findings demonstrated that for a given surface roughness height, an increase in the stratification strength of the free stream resulted in a reduction in the height of the CNBL. Conversely, for a fixed free-stream stratification strength, an increase in the surface roughness height led to an increase in the CNBL height. 

In \cite{Liu_et_al_2021}, the LES of CNBL flows were studied with a Geostrophic wind of magnitude $G=12$ m/s. In this study, the surface roughness height was fixed to $z_0=10^{-4}$ m while the Coriolis frequencies were varied representing the ABL flows in low and high-latitude regions. Three free-stream stratification strengths were considered such that the parameter $\mu_N$ ranged between 42 to 1350. In summary,  \cite{Liu_et_al_2021} performed twenty-four LES simulations of the CNBL flow. Table \ref{tab:data_CNBL} lists the values of the non-dimensional numbers $\mu_N$, $Ro$ and the dimensional estimates of $h,u_*,\alpha_0$ for these twenty-four cases.  \cite{Liu_et_al_2021} revisited the GDL model proposed in \cite{Zilitinkevich_Esau_2005} and provided updated model coefficients for predicting $u_*, \alpha_0$ for the CNBL flows. Consistent with the findings of \cite{Abkar_Porteagel_2013}, their results demonstrated that for a given surface roughness height, the CNBL depth gets shallower as the free stream stratification strength is increased.  \cite{Liu_et_al_2021} also showed that the CNBL height becomes smaller in high latitude regions characterized by an increase in the magnitude of the Coriolis frequency. 

For our model validation, we also consider another six sets of LES performed in \cite{Liu_et_al_2021_PRL} and \cite{Liu_Stevens_2022_PNAS}. These studies performed the LES  to predict the structure of the CNBL flows driven by a Geostrophic wind of magnitude $G=12$ m/s. \cite{Liu_et_al_2021_PRL} proposed an analytical model for the resultant wind velocity profile while in \cite{Liu_Stevens_2022_PNAS}, both the streamwise and spanwise components of the velocity profiles were obtained. Although the dimensional values of $h,u_*,\alpha_0$ for these cases are not explicitly reported in the references, we estimate these quantities using the GDL model described in \cite{Liu_Stevens_2022_PNAS} for the given values of $\mu_N$ and $Ro$. The corresponding values of $\mu_N$, $Ro$, $u_*$, $h$ and $\alpha_0$ for these six simulations are listed in  Table \ref{tab:data_CNBL}.

\begin{longtable}{cccccc}

\hline
Description &$\mu_N$ & $Ro$ & $h (m)$ & $u_* (m/s)$ & $\alpha_0 (^\circ)$ \\ 
\hline
&&&&&\\
\multirow{1}{*}{Data 1}  & 58 & $4.33\times 10^4$ & 793 & 0.4332 & 18.49 \\ 
    & 58 & $3.61\times 10^5$ & 661 & 0.3606 & 14.88 \\ 
    & 180 & $4.11\times 10^4$ & 433 & 0.4105 & 27.75 \\ 
    & 180 & $3.48\times10^5$ & 375 & 0.3477 & 23 \\
 & &&&&\\
    \hline
  & &&&&\\
   \multirow{1}{*}{Data 2}   & 42 & $2.3\times 10^7$ & 482 & 0.3084 & 11.88 \\ 
   & 72 & $2.3\times 10^7$ & 396 & 0.3084 & 13.02 \\ 
   & 125 & $2.3\times 10^7$ & 312 & 0.3096 & 14.98 \\ 
   & 45 & $2.4\times10^7$ & 507 & 0.3072 & 11.93 \\ 
   & 78 & $2.4\times 10^7$ & 413 & 0.3072 & 13.12 \\ 
   & 136 & $2.4\times 10^7$ & 326 & 0.3072 & 15.16 \\ 
   & 51 & $2.8\times 10^7$ & 552 & 0.3048 & 11.92 \\ 
   & 89 & $2.8\times 10^7$ & 438 & 0.3048 & 13.45 \\ 
   & 154 & $2.8\times 10^7$ & 350 & 0.3048 & 15.41 \\ 
   & 61 & $3.3\times 10^7$ & 602 & 0.3024 & 12.13 \\ 
   & 106 & $3.3\times 10^7$ & 477 & 0.3036 & 13.79 \\ 
   & 183 & $3.3\times 10^7$ & 375 & 0.3024 & 16.20 \\ 
   & 78 & $4.0\times 10^7$ & 692 & 0.2988 & 12.46 \\ 
   & 136 & $4.0\times 10^7$ & 533 & 0.3000 & 14.64 \\ 
   & 235 & $4.0\times 10^7$ & 417 & 0.2988 & 17.38 \\ 
   & 115 & $5.9\times 10^7$ & 835 & 0.2940 & 13.31 \\ 
   & 199 & $5.9\times 10^7$ & 636 & 0.2928 & 16.02 \\ 
   & 344 & $5.7\times 10^7$ & 485 & 0.2868 & 19.26 \\ 
   & 226 & $1.1\times 10^8$ & 1109 & 0.2820 & 16.12 \\ 
   & 391 & $1.1\times 10^8$ & 826 & 0.2736 & 20.02 \\ 
   & 678 & $1.0\times 10^8$ & 613 & 0.2604 & 24.88 \\ 
   & 557 & $1.5\times 10^8$ & 942 & 0.2628 & 22.89 \\ 
   & 965 & $1.4\times 10^8$ & 698 & 0.2472 & 28.02 \\ 
   & 1350 & $1.8\times 10^8$ & 778 & 0.2328 & 31.31 \\
   & &&&&\\
   \hline
    & &&&&\\
 \multirow{1}{*}{Data 3}  & 51.2 
   & $2.70\!\times\!10^7$ & 383 & 0.3135 & 11.96 \\ 
     & 88.7 & $4.50\!\times\!10^4$  & 861 & 0.5166 & 22.26 \\ 
   & 88.7 & $3.70\!\times\!10^5$ & 713 & 0.4278 & 18.28 \\ 
   & 88.7 & $3.20\!\times\!10^6$ & 604 & 0.3626 & 15.42 \\
   & 88.7 & $2.70\!\times\!10^7$ & 525 & 0.3147 & 13.34 \\ 
   & 153.6 & $2.70\!\times\!10^7$ & 703 & 0.3135 & 15.67\\
   & &&&&\\
    \hline
    \caption{Table lists the CNBL LES data from \cite{Abkar_Porteagel_2013} (Data 1), \cite{Liu_et_al_2021} (Data 2),  \cite{Liu_et_al_2021_PRL} (Data 3).}
    \label{tab:data_CNBL}
\end{longtable}

\subsection{Additional LES data including SBL flows\label{Sec:SBL_data}}

The LES data in Table \ref{tab:data_CNBL} consists only of CNBL cases. We perform additional LES simulations including SBL flows. We use the 
\href{https://lesgo.me.jhu.edu}{LESGO} 
solver that has seen many applications for ABL simulations
\citep{Albertson_1999,bouzeid2005,kumar2006large,sescu2015large,calaf_et_al_2010,stevens2014,stevens_et_al_JRSE_2014,sescu2014,Abkar_et_al_2018,shapiro2020}.

We discuss the governing equations and numerical methodology of the LESGO solver in Section \ref{gov_eqn_nm_sec}. We then present the simulation setup in Section \ref{sim_setup_sec}.

\subsubsection{Governing equations and numerical method \label{gov_eqn_nm_sec}}
LESGO solves the filtered Navier-Stokes equations with a buoyancy force term approximated using the Boussinesq approximation and the scalar potential temperature transport equation: 

\begin{align*}
    \frac{\partial \tilde{u}_i}{\partial x_i}&=0,\numberthis\label{continuity_LES}\\
    \frac{\partial \tilde{u}_i}{\partial t}+\tilde{u}_j\left(\frac{\partial \tilde{u}_i}{\partial x_j}-\frac{\partial \tilde{u}_j}{\partial x_i}\right)&=-\frac{1}{\rho_0}\frac{\partial p_\infty}{\partial x_i} -\frac{\partial \tilde{p}}{\partial x_i}+\frac{g}{\tilde{\theta}_0}(\tilde{\theta}-\tilde{\theta}_0)\delta_{i3}-\frac{\partial\tau_{ij}}{\partial x_j}\\
    &\mspace{20mu}-f_c\tilde{u} \ \delta_{i2}+f_c\tilde{v} \ \delta_{i1},\numberthis\label{momentum_LES}\\
    \frac{\partial\tilde{\theta}}{\partial t}+\tilde{u}_j\frac{\partial \tilde{\theta}}{\partial x_j}&=-\frac{\partial \Pi_j}{\partial x_j},\numberthis\label{theta_LES}
\end{align*}
where the tilde ($\tilde{\cdot}$) represents a spatial filtering operation such that $\tilde{u}_i=(\tilde{u},\tilde{v},\tilde{w})$ are the filtered velocity components in the streamwise, lateral and vertical directions, respectively, and $\tilde{\theta}$ is the filtered potential temperature. The term $\tau_{ij}=\sigma_{ij}-(1/3)\sigma_{kk}\delta_{ij}$ is the deviatoric part of the Sub-Grid Scale (SGS) stress tensor $\sigma_{ij}=\widetilde{u_i u_j}-\tilde{u}_i\tilde{u}_j$. The quantity $\tilde{p}=\tilde{p}_*/\rho_0+(1/3)\sigma_{kk}+(1/2)\tilde{u}_j\tilde{u}_j$ is the modified pressure, where the actual pressure $\tilde{p}_*$ divided by the ambient density $\rho_0$ is augmented with the trace of the SGS stress tensor and the kinematic pressure arising from writing the non-linear terms in rotational form. The $\delta_{ij}$ in the momentum equation \eqref{momentum_LES} is the Kronecker delta function determining the direction of the buoyancy and Coriolis forces. In the buoyancy term, $g=9.81 \ \text{m}/\text{s}^2$ is the gravitational acceleration, $\tilde{\theta}_0$ is the reference potential temperature scale. The Coriolis frequency in the Coriolis force term is defined as $f_c=2\Omega\sin\phi$, where $\phi$ is the latitude angle of a given region. In the potential temperature equation \eqref{theta_LES}, the term $\Pi_j=\widetilde{u_j \theta}-\tilde{u}_j\tilde{\theta}$ is the SGS heat flux. The SGS terms $\tau_{ij}$ and ${\Pi}_{j}$ are modeled as

\begin{align}
\tau_{ij}&=-2\nu^{\text{SGS}}_T\tilde{S}_{ij}, \mspace{50mu}
    \Pi_j=-\kappa^{\text{SGS}}_T\partial\tilde{\theta}/{\partial x_j},\label{SGS_model}
\end{align}
where $\tilde{S}_{ij}=(1/2)(\partial \tilde{u}_i/\partial x_j+\partial \tilde{u}_j/\partial x_i)$ is the symmetric part of the velocity gradient tensor, $\nu^{\text{SGS}}_T$ and $\kappa^{\text{SGS}}_T$ are the SGS momentum and heat diffusivities, respectively. The SGS Prandtl number 
$(Pr_{\text{SGS}})$
relates the two diffusivities, which are evaluated as

\begin{align}
    \nu_T^{\text{SGS}}&=(C_s\tilde{\Delta})^2\sqrt{\tilde{S}_{ij}\tilde{S}_{ij}}\label{nuT_SGS},\mspace{50mu}
    \kappa_T^{\text{SGS}}=Pr_{\text{SGS}}^{-1}\nu_T^{SGS}. 
\end{align}
Here, $C_s$ is the Smagorinsky model coefficient, and $\tilde{\Delta}=(\Delta x\Delta y \Delta z)^{1/3}$ is the effective grid spacing or filter width. The model coefficient $C_s$ is evaluated using the Lagrangian dynamic scale-dependent model \citep{bouzeid2005}.

To represent an ABL that is driven by a Geostrophic wind, we apply a mean pressure gradient $-(1/\rho_0)\partial p_\infty/\partial x_i$ in equation \eqref{momentum_LES}, which is determined from the Geostrophic balance:

\begin{align}
    {\rho_0}^{-1}{\partial p_\infty}/{\partial x}= f_c V_g,\quad {\rho_0}^{-1}{\partial p_\infty}/{\partial y}=-f_c U_g.\label{dpdx_ext}
\end{align}

Here, we set the Geostrophic wind magnitude $G=\sqrt{U_g^2+V_g^2}$ while the direction of the wind $\alpha$ (such that $U_g=G\cos\alpha,\ V_g=G\sin\alpha$) is controlled by a Proportional-Integral (PI) controller designed to impose a desired mean velocity orientation at a particular height \citep{sescu2014,Narasimhan_et_al_2022}. For future wind energy applications, we selected that the mean velocity is aligned in the streamwise direction at a height of 100 m, but simulation results are rotated in order to obtain zero spanwise mean velocity at the ground after the simulation is completed to be consistent with the model formulation. 

The code employs a pseudo-spectral technique to discretize the streamwise and spanwise directions. For discretizing the wall-normal direction, a second-order 
central finite difference method is utilized. To advance in time, the code employs the second-order accurate Adams-Bashforth scheme. In order to reduce the effects of streamwise periodicity, a shifted periodic boundary condition is employed \citep{muntersetal2016}. 

The effect of atmospheric stability is incorporated in the  boundary condition by evaluating the surface momentum fluxes utilizing the MOST expression for the mean velocity. Assuming the first grid point is within the ASL region, the corresponding surface momentum flux $\tau_w$ is given by

\begin{align}
    \tau_w=-\left(\frac{\tilde{u}_r\kappa}{\ln(z_1/z_0)-\Psi_{\text{m}}(z_1/L_s)+\Psi_{\text{m}}(z_0/L_s)}\right)^2,\label{tau_wall_LES}
\end{align}

where $\tilde{u}_r=\sqrt{\tilde{u}^2+\tilde{v}^2}$ is the resultant horizontal velocity at the first grid point $z=z_1=\Delta z/2$, $\kappa=0.41$ is the Von Karman constant, and $z_0$  is the surface roughness height, while $L_s$ is the Monin-Obukhov length. From equation \eqref{tau_wall_LES}, the surface momentum flux components are evaluated as:

\begin{align}
    \tau_{i,3|w}=\tau_w\times ({\tilde{u}_i}/{\tilde{u}_r}), \ i=1,2.\label{tau_wall_i3_LES}
\end{align}

These  are applied as a boundary condition at the bottom boundary, while a stress-free boundary condition is imposed on the top boundary.

The $L_s$ in equation \eqref{tau_wall_LES} is evaluated in LES as follows

\begin{align}
    L_s&={u_*^2\tilde{\theta}_1}/{(\kappa g T_*)}, \label{Ls_LES}
\end{align}

where the friction velocity is evaluated  as $u_*=\sqrt{|\tau_w|}$, $\tilde{\theta}_1$ is the potential temperature at the first grid point and $T_*=-Q_0/u_*$ is a temperature scale that represents the ratio of the surface heat flux and the friction velocity. For the SBL flows, $Q_0<0$ and correspondingly $T_*>0$ resulting in $L_s>0$. The  model for the potential temperature provides  $T_*$ according to

\begin{align}
    T_*=-\frac{Q_0}{u_*}&=\frac{\kappa [\tilde{\theta_1}-\tilde{\theta}_s]}{\ln(z_1/z_{0s})-\Psi_\text{h}(z_1/L_s)+\Psi_\text{h}(z_{0s}/L_s)}. \label{T_start_LES}
\end{align}

Here $\tilde{\theta}_s$ is the surface potential temperature and $z_{0s}$ is the surface roughness height for the potential temperature which is taken to be $z_{0s}=0.1 z_0$. To simulate an SBL flow, we decrease the surface temperature by specifying a constant cooling rate $C_r$ ($C_r<0$). We use this $C_r$ to evaluate $\tilde{\theta}_s(t)$ at a given  time step using $\tilde{\theta}_s(t)=\tilde{\theta}_s(t-\Delta t)+C_r\Delta t$.

The stability functions $\Psi_{\text{m}}(\zeta=z/L_s)$ and $\Psi_{\text{h}}(\zeta=z/L_s)$ in equations \eqref{tau_wall_LES} and \eqref{T_start_LES}, respectively, are obtained from \cite{Chenge_Brutsaert_2005} given by

\begin{align}
        \Psi_{\text{m}/\text{h}}(\zeta=z/L_s)&=
        -a_{\text{m}/\text{h}}\ln\left[\zeta+(1+\zeta^{b_{\text{m}/\text{h}}})^{b_{\text{m}/\text{h}}^{-1}}\right], \label{psimh_stable}
\end{align}

where the constants are $a_{\text{m}}=6.1$, $b_{\text{m}}=2.5$, $a_{\text{h}}=5.3$, $b_{\text{h}}=1.1$.

In order to dampen the gravity waves in the computational domain, a sponge (or Rayleigh damping) layer is used at the top boundary. This is a wave-absorbing layer spanning 500 meters from the top boundary. Within this layer, a body force with a cosine profile for its damping coefficient is employed to mitigate the reflection of gravity waves. \citep{allaerts_meyers_2017,Durran_Klemp_1983}.

\subsubsection{Simulation setup \label{sim_setup_sec}}

The LES for the current study is performed in a computational domain of size $L_x\times L_y\times L_z=3.75 \ \text{km} \times 1.5 \ \text{km} \times 2\ \text{km}$. The streamwise, spanwise, and wall-normal directions are discretized using $N_x\times N_y\times N_z=360\times 144\times 432$ grid points. The resulting grid resolution is $\Delta x\times\Delta y\times \Delta z=10.4 \ \text{m}\times 10.4 \ \text{m} \times 4.6 \ \text{m}$. We set the Geostrophic wind magnitude as $G=15$ m/s, Coriolis frequency $f_c=10^{-4}$ s$^{-1}$, surface roughness height $z_0=0.1$ m, SGS Prandtl number $Pr^{\text{SGS}}=1$.

\begin{longtable}{ccccccc}
		\hline
		&&&&&&\\
            Case & $\mu$ & $Ro$ &$C_r$ (K/hr)& $h$ (m)&  $u_*$ (m/s)&  $\alpha_0^{\circ}$\\
		&&&&&&\\ 
            \hline
            &&&&&&\\
		CNBL& 0 &  $6.02\times 10^4$ & 0 & 1157& 0.60 &21\\
            SBL-1& 5.62 & $5.93\times 10^4$&-0.03 & 1032&  0.59 &24\\
            SBL-2& 20.59 & $5.25\times 10^4$&-0.125 & 662&  0.53 &28\\
            SBL-3& 39.84 & $4.58\times 10^4$&-0.25 & 463& 0.46 &32\\
            SBL-4& 59.25 & $4.12\times 10^4$&-0.375 & 361&0.41 &35\\
            SBL-5& 78.35 & $3.84\times 10^4$&-0.5 & 306&0.38 &38\\
            SBL-6& 148.49& $3.39\times 10^4$&-1 & 218& 0.34 &41\\
            &&&&&\\
            \hline
            &&&&&\\
 \caption{\label{tab:LESGO} Description of LES of CNBL and SBL cases with corresponding values of stability parameter $\mu$, friction Rossby number $Ro$,  Cooling rate $C_r$ (K/hr), ABL height $h$ (m), friction velocity $u_*$ (m/s), and cross-isobaric angle $\alpha_0^\circ$. For all simulations, $G=15$ m/s, $z_0=0.1$ m, $\Theta_0=265$ K. The Coriolis frequency is $f_c=10^{-4}$ s$^{-1}$ and the free-stream Brunt-V\"ais\"al\"a frequency is $N_\infty=6.1\times 10^{-3}$ s$^{-1}$ which gives the Zilitinkevich number $\mu_N=61$. }
\end{longtable}

For both the CNBL and SBL simulations, the velocity fields are initialized with a log-law velocity profile  superimposed with a zero-mean white noise within the first 100 m from the surface to initiate turbulence. A description of the respective setups of the initial potential temperature profile for the CNBL and SBL simulations is as follows.

We first perform the LES of a CNBL flow using an initial linear potential temperature profile $\Theta(z)=\Theta_0+\gamma_\Theta z$ with $\Theta_0=265$ K, and $\gamma_\Theta=0.001$ K/m. The simulation reaches a quasi-stationary state where the boundary layer depth grows to 1157 m. The resulting temperature profile has a capping inversion layer at the ABL depth which separates the neutral boundary layer region from the stably stratified Geostrophic region. Under quasi-steady conditions, the potential temperature within the CNBL region stayed around 265.58 K. 

For the LES of SBL flows, we initialize potential temperature using the CNBL's quasi-steady potential temperature profile. We then decrease the magnitude of the surface potential temperature $\tilde{\theta}_s$  by applying different cooling rates $C_r=[-0.03,-0.125,-0.25,-0.375,-0.5,-1]$ K/hr to induce stable stratification. The simulations are run until reaching  a quasi-steady state. We then perform time and planar averaging to obtain the vertical profiles of the turbulent stresses and mean wind velocities. The resultant values of $h, u_*, \alpha_0$ are listed in  Table \ref{tab:LESGO}. These estimates are used for the validation of the new GDL model. Since the PI controller maintains a streamwise aligned mean flow at $z=100$ m, the $\alpha_0$ values reported in Table \ref{tab:LESGO} are obtained by geometrically rotating the mean velocity profiles such that wind veer is zero at the first grid point. This is done to be consistent with the GDL model derivation using a coordinate system in which there is no wind veer within the ASL region.

In the following section, we use these forty-one  LES cases covering a range of atmospheric conditions for validation of the comprehensive ABL wind model discussed in section \S\ref{Sec:2}.

\section{Results and discussion}\label{Sec:Results}

In this section, we validate the ABL wind model by comparing the model predictions with the corresponding LES cases. We first compare our new GDL  approach in section \S\ref{Sec:GDL_comparison}. We then compare the ABL wind velocity profile predictions with the LES in section \S\ref{Sec:velocity_comparison}.

\subsection{GDL: LES vs Model}\label{Sec:GDL_comparison}

We determined the ABL height $h$, friction velocity $u_*$, and the cross-isobaric angle $\alpha_0$ from the new GDL model described in \S \ref{Sec:2} using the iterative procedure described in the \ref{Sec:appendix}. For each of the forty-one LES cases described in \S\ref{Sec:LES}, we use the inputs ($G$, $z_0$, $f_c$, $C_r$, $N_\infty$) that are known from the LES  and  predict the unknown values $h$, $u_*$, and $\alpha_0$ using the new GDL model.

\begin{figure}[t]
    \centering
\includegraphics{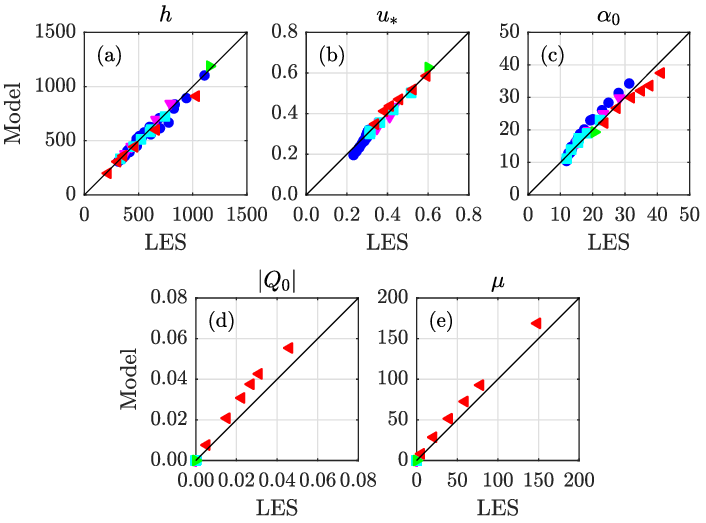}
    \caption{GDL predictions of (a) ABL height $h$ (in meters), (b) friction velocity $u_*$ (m/s), (c) cross-isobaric angle $\alpha_0$ (in degrees), (d) surface cooling flux magnitude $Q_0=C_r h$ (Eq. \ref{Q0_model}) in (K m/s), (e) stability parameter $\mu$ (Eq. \ref{mu_model}) compared against the LES of CNBL cases from \cite{Liu_et_al_2021} ($\protect\Liuetala$), \cite{Abkar_Porteagel_2013} ($\protect\AP$), \cite{Liu_Stevens_2022_PNAS} ($\protect\LS$), CNBL LES ($\protect\CNBLLESGO$) \& LES of SBL ($\protect\SBLLESGO$) from current work.}
    \label{GDL_plot}
\end{figure}

The GDL model predictions are compared with the LES in Figures \ref{GDL_plot}(a), (b), and (c). The colored markers represent the forty-one LES cases. The markers ($\protect\CNBLLESGO$) and ($\protect\SBLLESGO$) represent the CNBL and SBL cases, respectively, from the LES cases in Table \ref{tab:LESGO} while the rest of the markers represent the CNBL data from previous studies (Table \ref{tab:data_CNBL}). 
The GDL predictions for $h,u_*,\alpha_0$ lie close to the solid black line signifying excellent agreement with the LES. The comparison of the model estimates for the surface cooling flux $Q_0=C_r h$ (Eq. \ref{Q0_model}) and the stability parameter $\mu$ (Eq. \ref{mu_model}) are shown in Figure \ref{GDL_plot}(d) and (e). While $Q_0=\mu=0$ for the CNBL cases, the non-zero values corresponding to the SBL cases show good agreement with the LES.

These results suggest that the newly proposed  GDL formulation is able to predict  ABL depth $h$, friction velocity $u_*$, and cross-isobaric angle $\alpha_0$ correctly for the different flow and thermal stability conditions for the data in Tables \ref{tab:data_CNBL} and \ref{tab:LESGO}.  

\subsection{Velocity profiles: LES vs Model\label{Sec:velocity_comparison}}

In this section, we compare the ABL wind velocity profiles from the model with the LES for the cases described in Table \ref{tab:LESGO}.

\begin{figure}[t]
    \centering
\includegraphics[scale=1]{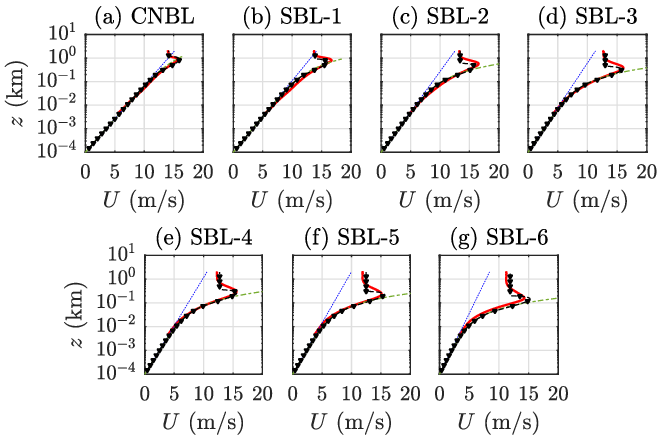}
    \caption{Comparing model and LES velocity profiles. Model predictions of $U(z)$ ($\protect\Umodel$, Eq. \ref{U_eqn_model}) compared against Log-law profile ($\protect\bluedashedline$), the MOST prediction with CNBL correction from Eq. \eqref{U_inner} ($\protect\UMOST$)  
    and profiles from LES ($\protect\redline$) of the (a) CNBL, (b) SBL-1, (c) SBL-2, (d) SBL-3, (e) SBL-4, (f) SBL-5, (g) SBL-6. }
    \label{UV_log_plot}
\end{figure}

We first show the comparison of the vertical variation of the streamwise velocity in Figures \ref{UV_log_plot}(a)-(g). The plot is made using a logarithmic scale for the vertical direction. The predictions are made for the LES cases in Table \ref{tab:LESGO}. Plot \ref{UV_log_plot}(a) corresponds to the CNBL case while the Figures \ref{UV_log_plot}(b)-(g) are from the SBL cases where the cooling rate is progressively  increased from -0.03 K/hr to -1 K/hr. The solid red lines in these plots are from the LES while the black dashed lines with a triangular marker are the model predictions from Eq. \eqref{U_eqn_model}. Following the model assumption of zero wind veer within the ASL, the LES estimates of $U(z)$ are obtained after performing a geometric rotation satisfying this zero wind veer condition. The blue dashed line represents the log-law profile. The green dash-dotted line is the model prediction from Eq. \eqref{U_inner}, the MOST prediction including the CNBL correction.
All the streamwise velocity curves follow the log law close to the ground within the constant stress region of the ASL. Upon increasing the cooling rate, the stable stratification causes deviation of the velocity profile from the log law behavior. This stability-affected region is in the $z$-independent stratification layer as it matches the shape from the MOST with the CNBL correction. Above the ASL lies the Ekman layer, where the MOST predictions would   significantly deviate from the LES and also not satisfy the Geostrophic condition. Whereas the ABL wind model Eq. (\ref{U_eqn_model}) captures the low-level jet profile  where it reaches a peak velocity at some height and then decreases to the Geostrophic wind velocity at the ABL depth. The flow velocity in this Geostrophic region is set to $U_g$ predicted from the GDL model, again in very good agreement with the LES. The discussion shows that the newly proposed ABL wind model is able to capture the streamwise velocity structure of the entire ABL spanning across different thermal stability conditions. 

\begin{figure}[t]
    \centering
\includegraphics[scale=1]{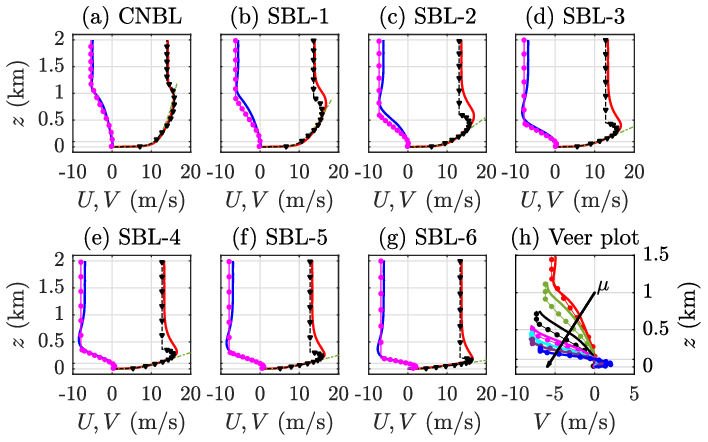}
    \caption{Model predictions of $U(z)$ ($\protect\Umodel$, Eq. \ref{U_eqn_model}) and $V(z)$ ($\protect\Vmodeld$, Eq. \ref{V_eqn_model}) compared against the MOST model ($\protect\UMOST$, Eq. \ref{U_inner}) and LES ($U(z)$ - $\protect\redline$, $V(z)$ - $\protect\blueline$) from (a) CNBL, (b) SBL-1, (c) SBL-2, (d) SBL-3, (e) SBL-4, (f) SBL-5, (g) SBL-6. The veer plot in (h) shows how stronger stable stratification (increasing $\mu$) causes a decrease in the ABL height, resulting in a strong wind veer flow. Figure (h) consists of the analytical wind veer estimates that are represented by a circle marker on a thin solid line and different colors represent CNBL ($\protect\Vmodela$), SBL-1 ($\protect\Vmodelb$), SBL-2 ($\protect\Vmodelc$), SBL-3 ($\protect\Vmodeld$), SBL-4 ($\protect\Vmodele$), SBL-5 ($\protect\Vmodelf$), SBL-6 ($\protect\Vmodelg$). The corresponding LES estimates of $V(z)$ in (h) are plotted as solid lines without the circle marker using the same color as analytical estimates.}
    \label{UV_plot}
\end{figure}

The predictions of both the streamwise and spanwise velocity components from Eqs. \eqref{U_eqn_model}, \eqref{V_eqn_model} are compared against the LES in Figures \ref{UV_plot}(a)-(h) in linear units, and without rotating the velocities to zero veer at the ground but applying the model with zero veer at 100m height, as in the LES. The horizontal line just above the origin in the Figures \ref{UV_plot}(a)-(h) represents this 100 m height where we see that $V(z=100 \ \text{m})=0$ in all these figures.
The solid red and blue  lines are the streamwise and spanwise velocity components from the LES, respectively. The black lines with the triangular markers represent the model predictions for $U(z)$. Similarly, the magenta lines with circle markers are the model predictions of $V(z)$. The MOST prediction with the CNBL correction from Eq. \eqref{U_inner} is also plotted as a green dash-dotted line.

The analytical velocity profiles from Eqs. \eqref{U_eqn_model}, \eqref{V_eqn_model} show  good overall agreement with the LES. More specifically, the CNBL prediction is in excellent agreement with the LES, while slight deviations exist for the SBL cases near the ABL height. These differences are attributed to non-zero turbulent stresses above the SBL region. In developing the analytical model, we assumed the turbulent stresses were zero above the ABL depth. However, from the LES, we see non-zero residual stresses exist above the boundary layer,  due to inertial oscillations.  This non-zero residual stress region is evident in Figure \ref{stress_plot}(b) where we show the vertical profiles of the normalized turbulent stress $\T{T}_{yz}(\hat{\xi})$.

The ABL wind model is able to predict the formation of a low-level jet for different stability conditions. The model also captures the decrease in the SBL height as we increase the cooling rate.
Most importantly, the model also predicts the wind veer profiles while theories like the MOST do not model the wind direction. The wind veer velocity profiles within the ABL region for all the cases are plotted together in Figure \ref{UV_plot}(h). The plot clearly shows the effect of increasing the cooling rate. Under strong stable stratification, the wind veer strength is intensified closer to the ground causing a significant cross-wind flow. The ABL wind model is able to capture these intense wind veer profiles that can be used as an input in Gaussian models for modeling wind turbine wakes \citep{bastankhah2014} or pollution puffs \citep{Zannetti1990} in the ABL. Another important aspect of the model is the accurate prediction of $u_*$ and Geostrophic velocities $U_g, V_g$ from the self-consistent drag law model. This new model enabled the prediction of the ABL wind velocities for the entire domain that is close to LES (Figure \ref{UV_plot}). 

In summary, the above comparisons demonstrate the capability of the new ABL wind model. The model can successfully predict the ABL wind velocity profiles for the entire domain across both conventionally neutral and stable atmospheric conditions.

\section{Conclusions}\label{Sec:conclusion}

In this study, we developed an analytical model to predict the steady-state mean velocity profiles for thermally stratified and conventionally neutral ABL flows.  We showed the turbulent stress components are approximately self-similar for the CNBL and SBL flows. We directly model these turbulent stresses using analytical formulations representing the self-similarity in the stresses using \cite{Nieuwstadt_1984}'s 3/2 power law, and an inner layer consistent with MOST-based modeling. These stress profiles were incorporated into the Ekman mean momentum equations to predict the ABL velocity components.  
Furthermore, we derived a self-consistent Geostrophic drag law model by matching the streamwise velocity in the inner and outer layer regions at a specific height and evaluating the spanwise velocity at the surface roughness height. We used an LES-based equilibrium ABL height model which predicts the ABL depth for the different types of boundary layer spanning neutral and stable boundary layer flows. 

The effects of thermal stratification are characterized by the stability parameters $\mu_N$ and $\mu$ which influenced the model expressions for turbulent stresses, velocity components, and the GDL. We assumed the parameter Zilitinkevich number $\mu_N$ is known and modeled the stability parameter $\mu$ by representing the surface cooling flux as the product of the cooling rate and  ABL depth.  We used this modeled $\mu$ in the analytical expressions for the velocities, turbulent stresses, and the GDL. 

To validate our model, we compared our predictions with corresponding values obtained from  LES from literature and new cases run specifically for SBL flows, and demonstrated a good agreement between them. Additionally, we compared model predicted wind velocity profiles with LES data, revealing that the new model accurately predicts the ABL mean velocity distribution under different atmospheric stability conditions. In conclusion, the new analytical model provides reliable predictions of ABL velocity profiles, capturing the MOST velocity profiles within the surface layer and the Ekman spiral structure within the Ekman layer, which eventually merges with the Geostrophic wind above the ABL.  

Future work aims to extend the approach to also describe convective boundary layers and perhaps allow to capture unsteady effects such as during a daily cycle. Development of further model refinements including other effects such as momentum exchanges due to canopies \citep{patton2016atmospheric} or wind farms \citep{calaf_et_al_2010} is also of significant interest.

{\bf Acknowledgements:}  This work is supported by the National Science Foundation (grants CBET-1949778 and CMMI-2034111).

\newpage
\myappendix

\section{: Summary of ABL model and iterative solution method}\label{Sec:appendix}

In this Appendix section, we summarize the iterative method to solve the coupled equations for the proposed ABL model discussed in section \S\ref{Sec:2}.
    \begin{enumerate}
        \item Input values: $N_\infty,f_c,G,z_0,C_r$  (assuming $f_c>0$, northern hemisphere).
        \item Iterative calculation of $h,u_*,\alpha_0$:
        \begin{enumerate}
            \item Assume initial values for $h^0,u_*^0$ and set $h^n=h^0,u_*^n=u_*^0$.
            \item Compute $u_*^{n+1}$ using Eq. \eqref{u_tau_model}:
            
            \begin{align*}
            u_*^{n+1}&=\frac{\kappa G}{\sqrt{[\ln \left(Ro^n\right)-A^n]^2+(B^n)^2}}, \,\, {\rm where} \numberthis\label{u_tau_iter1}
            \end{align*}
            
            \begin{align*}
            A^n&=-\ln c_m \hat{h}^n-\kappa\Biggl[(5\mu^n+0.3\mu_N)(c_m\hat{h}^n-\hat{\xi}_0^n)\\
            &\mspace{20mu}
            +g^\prime(c_m\hat{h}^n)\left(1-c_m\right)^{3/2}
            -g(c_m\hat{h}^n)\frac{3}{2\hat{h}^n}\sqrt{1-c_m}\Biggr],\\
            B^n&=\frac{3\kappa}{2\hat{h}^n}, \,\,\, {\rm and} 
            \end{align*} 
            
            \begin{align*}
             g(c_m\hat{h}^n)&=M\left[1-e^{-\Gamma c_m}\right],\ g^\prime(c_m\hat{h}^n)=M\frac{\Gamma}{\hat{h}^n}e^{-\Gamma c_m}, \ M=1.43, \ \Gamma=1.2,\\
             \mu^n&=\frac{g\, (-C_r)}{u_*^n f_c^2 \, \Theta_0} \hat{h}^n, \ \,\,
             \hat{\xi}_0=\frac{z_0 f_c}{u_*^n}, \,\,\ \hat{h}^n=\frac{h^n f_c}{u_*^n}, \,\,  c_m=0.20.
            \end{align*}         
            
            \item Compute $h^{n+1}$ using $u_*^{n+1}$ from Eq. \eqref{u_tau_iter1} in Eq. \eqref{h_hat_mu_model}:
            
            \begin{align*}
            h^{n+1}&=\frac{u_*^{n+1}}{f_c}\left[\frac{1}{C_{TN}^2}+\frac{\mu_N}{C_{CN}^2}+\frac{1}{C_{NS}^2}\frac{(g/\theta_0)(-C_r)h^n }{ {u_*^{n+1}}^2 f_c}\right]^{-1/2},\\
            C_{TN}&=0.5, \ C_{CN}=1.6, \ C_{NS}=0.78, \ \mu_N=N_\infty/f_c.
            \end{align*}
            
            \item  Iterate till convergence to get final values of $u_*,h$. 
            \end{enumerate}
          \item  Then evaluate:
          
            \begin{align*}
                \hat{h}&=\frac{h f_c }{u_*}, \ \hat{\xi}_0=\frac{z_0  f_c }{u_*},\
                \mu=\frac{g \,(-C_r)}{u_* f_c^2\,\Theta_0}\hat{h}, \
                Ro=\frac{u_*}{z_0 f_c},
            \end{align*}
            
            as well as the converged values of $A$ and $B$.
            \item Evaluate $U_g,V_g$ using the GDL equations (Eqs. \ref{GDL_A}, \ref{GDL_B}, \ref{A_model_GDL}, \ref{B_model_GDL}):
            
            \begin{align*}
              U_g  =\frac{u_*}{\kappa} \left(\ln\left(Ro\right)-A \right), \ \,\,
             V_g = - \frac{u_*}{\kappa} B,
            \end{align*}        
              
            and obtain $\alpha_0=\tan^{-1}\left(\dfrac{V_g}{U_g}\right)$.
        \item Evaluate $U(z)$ and $V(z)$ with $\hat{\xi} = z f_c/u_*$:
        
        \begin{align*}
        U(z)&= 
        \begin{cases}
        u_*\left( -g^\prime(\hat{\xi})\left[1-\dfrac{\hat{\xi}}{\hat{h}}\right]^{3/2}+g(\hat{\xi})\dfrac{3}{2\hat{h}}\sqrt{1-\dfrac{\hat{\xi}}{\hat{h}}} \right) +  U_g
        &, \ \hat{\xi}\geq\hat{\xi}_m\\
        &\\
        u_*\left( \dfrac{1}{\kappa}\ln\dfrac{\hat{\xi}}{\hat{\xi}_0}+(5\mu+0.3\mu_N)(\hat{\xi}-\hat{\xi}_0)\right) &,\ \hat{\xi}\leq\hat{\xi}_m
        \end{cases}.
        \end{align*}   
        
        \begin{align*}
        V(z)&= u_* \left( \dfrac{g(\hat{\xi})g^\prime(\hat{\xi})}{\sqrt{1-g(\hat{\xi})^2}}\left[1-\dfrac{\hat{\xi}}{\hat{h}}\right]^{3/2}+\dfrac{3}{2\hat{h}}\sqrt{1-g(\hat{\xi})^2}\left[1-\dfrac{\hat{\xi}}{\hat{h}}\right]^{1/2}\right) + V_g  .
        \end{align*}
        
    \end{enumerate}
\newpage
\bibliographystyle{spbasic_updated.bst}     
\bibliography{sample_library.bib} 

\end{document}